%% file: HDSR_format.tex
\documentclass[12pt]{article}

\usepackage{scicite}
\usepackage{times}
\usepackage{booktabs}
\usepackage{graphicx}
\usepackage{hyperref}
\usepackage{color}
\usepackage{subcaption}
\usepackage{longtable}
\usepackage{enumerate}
\usepackage{mathtools}

\graphicspath{ {./Figures/} }
\def\age{\textrm{age}}

\newenvironment{sciabstract}{%
\begin{quote} \bf}
{\end{quote}}

\title{The {\it age\/} of secrecy and unfairness in recidivism prediction
}

\author
{Cynthia Rudin$^{\ast}$, Caroline Wang$^\dag$ Beau Coker$^\dag$\\
\\
\normalsize{Departments of Computer Science, Electrical and Computer Engineering, and Statistical Science,}\\
\normalsize{Duke University,}\\
\normalsize{LSRC Building, Durham, NC 27708, USA}\\
}

\date{}

\begin{document} 

% Double-space the manuscript.
\baselineskip18pt
%\baselineskip24pt

% Make the title.

\maketitle 

% Place your abstract within the special {sciabstract} environment.

\begin{sciabstract}
\footnote{$^\ast$To whom correspondence should be addressed; E-mail: cynthia@cs.duke.edu. Symbol $^\dag$ denotes equal contribution} 
In our current society, secret algorithms make important decisions about individuals. 
There has been substantial discussion about whether these algorithms are unfair to groups of individuals. While noble, this pursuit is complex and ultimately stagnating because there is no clear definition of fairness and competing definitions are largely incompatible. 
We argue that the focus on the question of fairness is misplaced, as these algorithms fail to meet a more important and yet readily obtainable goal: transparency. 
As a result, creators of secret algorithms can provide incomplete or misleading descriptions about how their models work, and various other kinds of errors can easily go unnoticed. By trying to partially reconstruct the COMPAS model --- a recidivism-risk scoring model used throughout the criminal justice system --- we show that it does not seem to depend linearly on the defendant's age, despite statements to the contrary by the model's creator. This observation has not been made before despite many recently published papers on this algorithm. Furthermore, by subtracting from COMPAS its (hypothesized) nonlinear age component, we show that COMPAS does not necessarily depend on race, contradicting ProPublica's analysis, which assumed linearity in age. In other words, faulty assumptions about a proprietary algorithm lead to faulty conclusions that go unchecked. Were the algorithm transparent in the first place, this likely would not have occurred.
We demonstrate other issues with definitions of fairness and lack of transparency in the context of COMPAS, including that a simple model based entirely on a defendant's age is as `unfair' as COMPAS by ProPublica's chosen definition. We find that there are many defendants with low risk score but long criminal histories, suggesting that data inconsistencies occur frequently in criminal justice databases. We argue that transparency satisfies a different notion of procedural fairness by providing both the defendants and the public with the opportunity to scrutinize the methodology and calculations behind risk scores for recidivism.
\end{sciabstract}

\input{HDSRcontent.tex}

% In setting up this template for *Science* papers, we've used both
% the \section* command and the \paragraph* command for topical
% divisions.  Which you use will of course depend on the type of paper
% you're writing.  Review Articles tend to have displayed headings, for
% which \section* is more appropriate; Research Articles, when they have
% formal topical divisions at all, tend to signal them with bold text
% that runs into the paragraph, for which \paragraph* is the right
% choice.  Either way, use the asterisk (*) modifier, as shown, to
% suppress numbering.

%% References
% Your references go at the end of the main text, and before the
% figures.  For this document we've used BibTeX, the .bib file
% scibib.bib, and the .bst file Science.bst.  The package scicite.sty
% was included to format the reference numbers according to *Science*
% style.

%BibTeX users: After compilation, comment out the following two lines and paste in
% the generated .bbl file. 
%\bibliographystyle{Science_v2}
%\bibliography{recidivism}

\end{document}

%% file: HDSRcontent.tex
\section{Introduction}
Secret algorithms control important decisions about individuals, such as judicial bail, parole, sentencing, lending decisions, credit scoring, marketing, and access to social services. These algorithms may not do what we think they do, and they may not do what we want. 

% Intro to fairness and COMPAS debate.
There have been numerous debates about fairness in the literature, mainly stemming from a flawed analysis by the ProPublica group \cite{LarsonMaKiAn16,AngwinLaMaKi16} of data from Broward County FL, claiming that the proprietary prediction model COMPAS (Correctional Offender Management Profiling for Alternative Sanctions) \cite{northpointe} is racially biased. COMPAS is used throughout the criminal justice system in the U.S., and its predictions have serious consequences in the lives of many people \cite{forbesunfairness,Corbett2016,pasc2018impact, netter2007using,lowenkamp2004understanding,gottfredson1996race,redding2009evidence,baradaran2013race,petersilia1987guideline,crow2008complexities,flores16}. The bottom line from these debates is that there is not a single correct definition of fairness and the fact that multiple types of fairness are incompatible.  We put aside typical fairness considerations for a moment to focus on more pressing issues.

% COMPAS problem 1: it's complicated
One issue with COMPAS is that it is \textit{complicated}. It is based on 137 variables \cite{compasquestionnaire} that are collected from a questionnaire. This is a serious problem because typographical or data entry errors, data integration errors, missing data, and other types of errors abound when relying on manually-entered data. Individuals with long criminal histories are sometimes given low COMPAS scores (which labels them as low risk), and vice versa. In the past, there have been documented cases where individuals have received incorrect COMPAS scores based on incorrect criminal history data \cite{nyt-computers-crim-justice,WexlerGlenn2017} and have had no mechanism to correct it after a decision was made based on that incorrect score. This problem has inadvertently occurred with other (even transparent) scoring systems, at least in one case leading to the release of a dangerous individual who committed a murder while on bail \cite{npr-bail:2017, Ho2017}. An error in a complicated model is much harder to find than an error in a simple model, and it not clear how many times typographical errors in complicated models have led to inappropriate releases that resulted in crimes, after decades of widespread use of these models. The question of whether calculation errors occur often in these models is of central importance to the present work.

% COMPAS problem 2: it's proprietary
A separate issue with COMPAS is that it is \textit{proprietary}, which means its calculations cannot be double-checked for individual cases, and its methodology cannot be verified. Furthermore, it is unclear how the data COMPAS collects contribute to its automated assessments. For instance, while some of the questions on the COMPAS questionnaire are the same as those in almost every risk score --- age, and number of past crimes committed --- other questions seem to be direct proxies for socioeconomic status, such as ``How hard is it for you to find a job ABOVE minimum wage compared to others?" It is not clear that such data should be collected or permitted for the purposes in which these risk scores are used.

% Consequences of using a proprietary model. Most of our results are referenced here.
% By nature, proprietary algorithms are more difficult to analyze and permit the algorithm's creators to provide incomplete or misleading descriptions about how they work.
Though creators of proprietary algorithms often provide descriptions of how their models work, by nature, it is difficult for third-parties to verify these descriptions. This may allow errors in documentation to propagate unchecked for years. By partially reconstructing COMPAS in Broward County, we show in Section \ref{sec:howcompasdependsonage} that COMPAS may depend nonlinearly on age, contradicting its stated methodology. As a result, ProPublica's conclusion that being African-American leads to a higher COMPAS score, even controlling for criminal history and sex, based on a logistic regression would be invalid because the linearity assumption is violated. While adding a nonlinear age term to the regression could mitigate the impact of this particular issue, it misses the larger point. Without transparency, incorrect high-level statements about a model (e.g., its dependence on age is linear) go unchecked, and this can have downstream consequences for independent analyses of a model.

While COMPAS depends heavily on age, we show in Sections \ref{sec:crimhist_general} through \ref{subsec:propublicarace} that it does not seem to depend in such a strong way on either criminal history or proxies for race. We discuss several possible reasons, but it is possible that COMPAS depends less on criminal history than we might expect. This leads to the possibility that COMPAS depends heavily on variables that we may not want it to depend on. 

 In Section \ref{sec:typos} we pinpoint many individuals whose COMPAS scores seem unusually low given their criminal histories. Since COMPAS is proprietary, we cannot fully determine whether these low scores are due to errors in calculation, data entry errors or errors from some other source (or even if they are errors at all). 

COMPAS' creator Northpointe disagreed with each of ProPublica's claims on racial bias \cite{northpointeresponse} based on their definition of fairness. Their rebuttal did not include arguments as to the type of fairness we consider here, and in particular, why it benefits the justice system to use a model that is complicated or proprietary.

% Complicated is unnecessary
% Why do we need complicated proprietary models?
Work in machine learning has shown that complicated, black-box, proprietary models are not necessary for recidivism risk assessment. Researchers have shown (on several datasets, including the data from Broward County)  that interpretable models are just as accurate as black box machine learning models for predicting recidivism \cite{ZengUsRu2017,tollenaar2013method,AngelinoLaAlSeRu17-kdd,angelino2018,RudinUs18,UstunRu19}. 
%and for many other applications \cite{Holte93,Freitas:2014ic,pitfall}. 
These simple models involve age and counts of past crimes, and indicate that younger people, and those with longer criminal histories, are more likely to reoffend. A judge could easily memorize the models within these works, and compute the risk assessments without even a calculator \cite{AngelinoLaAlSeRu17-kdd,angelino2018}. Despite this knowledge, complicated models are still being used.

Given that we do not need proprietary models, why we should allow proprietary models at all? The answer is the same as it is in any other application: by protecting intellectual property, we incentivize companies to perform research and development. Since COMPAS has been at the forefront of the fairness debate about modern machine learning methods, it is easy to forget that COMPAS is not one of these methods. It is a product of years of painstaking theoretical and empirical sociological study. For a company like Northpointe to invest the time and effort into creating such a model, it seems reasonable to afford the company intellectual property protections. However, as we discussed, machine learning methods --- either standard black-box or, better yet, recently-developed interpretable ones --- can predict equally well or better than bespoke models like COMPAS \cite{ZengUsRu2017,tollenaar2013method,AngelinoLaAlSeRu17-kdd,angelino2018}. For important applications like criminal justice, academics have always been willing to devote their time and energy. High-performing predictive models can therefore be created with no cost to the criminal justice system. 
%The advantages of a transparent model over a proprietary model are and always have been obvious, but 
Allowing proprietary models to incentivize model development is not necessary in the first place. %Like using a non-transparent, black-box model, using a proprietary model is simply no longer necessary because of methodological developments.

% Transparency as fairness
Neglecting to use transparent models has consequences. We provide two arguments for why transparency should be prioritized over other forms of fairness. First, no matter which technical definition of fairness one chooses, \textit{it is easier to debate the fairness of a transparent model than a proprietary model}. 
%We consider a simple recidivism model that uses only a defendant's age as a feature, which is widely considered to be an acceptable feature to use. We show this model is equally as unfair as COMPAS by ProPublica's definition, suggesting that \textit{any} model that depends on age will be unfair by this definition.
Transparent algorithms provide defendants and the public with imperative information about tools used for safety and justice, allowing a wider audience to  participate in the discussion of fairness.
Second, \textit{transparency constitutes its own type of procedural fairness} that should be seriously considered 
(see \cite{CoglianeseLe18} for a discussion). 
We argue that it is not fair that life-changing decisions are made with an error-prone system, without entitlement to a clear, verifiable, explanation. 

In Section \ref{sec:reverse_engineer}, we try to partially reconstruct COMPAS for Broward County and show how it is likely to be inconsistent with its official documentation; in Section \ref{sec:typos}, we identify a number of individuals with long criminal histories but low risk scores; and in Section \ref{sec:ageonly}, we describe transparency as a form of fairness. We consider the most transparent non-trivial predictive model we could find: age. Younger people tend to be at higher risk of recidivism. Our goal in this section is to modify the discussion of fairness to be through the lens of transparency. 

%%%%%%%%%%%%%%%%%%
\section{Reconstructing COMPAS}
\label{sec:reverse_engineer}
Even with our limited data, we may have been able to partially reconstruct parts of the COMPAS model as it is implemented in Broward County. We will next describe these attempts. Note that other groups besides ProPublica have attempted to explain COMPAS  \cite{caruana18,StevensonSl18}. One attempt \cite{StevensonSl18} uses only linear models (and our evidence suggests that COMPAS' dependence on age is nonlinear). Another attempt \cite{caruana18} modeled COMPAS with a generalized additive model, and we hypothesize that they had the correct model form, based on COMPAS' documentation. However, their analysis has similar issues to that of ProPublica or \cite{StevensonSl18}:
all of these groups attempted to explain COMPAS scores using ProPublica's limited set of features, but this type of analysis is not valid because there are many unmeasured features. COMPAS's actual dependence on the observable features may be totally different than what they report. For instance, if age is correlated with unmeasured survey data, their model would exploit this correlation to compensate for the missing survey data, leading to an incorrect estimate of how COMPAS depends on age.    
%rather than to actually determine what factors are important to COMPAS and how they contribute. 
Our approach instead attempts to isolate and subtract off the parts of COMPAS that we think can be explained from our data, using COMPAS' documentation to guide us. If our stated assumptions are correct, our analysis is valid even in the presence of many unmeasured features. The analysis still holds even if the measured and unmeasured features are correlated. Of course, ground truth cannot be made available (except to the designers of COMPAS), so our hypotheses cannot be verified.  

\subsection{COMPAS as described by its creator}
\label{sec:compas}

% I feel like a quick description of what COMPAS is (deciles vs raw score, general vs violent, subscales vs components of subscales and also what data we have will really help people read the paper.

There are two COMPAS recidivism risk assessments of interest: the general score and the violence score. These scores assess the risk that a defendant will commit a general or violent crime, and we use them to predict crime within the next two years. Each score is given as an integer between 1 and 10 but is based on a raw score that can take any value. Higher scores indicate higher risk. The raw score is computed by a formula and the final integer score is normalized based on a local population. We will therefore attempt to reconstruct the raw scores. 

To compute the COMPAS raw scores, Northpointe collects 137 variables from a questionnaire, computes a variety of \textit{subscales}, and finally linearly combines the subscales and two age variables --- the defendant's age at the time of the current offense and age at the time of the first offense --- to compute the raw risk scores. For example, using the equation exactly as written in the COMPAS documentation, the violent recidivism raw score is given by:
\begin{eqnarray*}
\lefteqn{\textrm{Violent Recidivism Risk Score}} \\
&=& \textrm{(age$*-$w)+(age-at-first-arrest$*-$w)+(history of violence $*$ w)} \\
&&+ \textrm{(vocation education $*$ w) + (history of noncompliance $*$ w),}
\end{eqnarray*}
where the variables not related to age are subscales and the weights ``w'' may be different for each variable. The notation ``age$*-$w'' would commonly indicate ``age times the negative of w." We have little knowledge of how the subscales depend on the questionnaire; the documentation states only which questionnaire items are used for which subscales. Table \ref{table:subscales} shows for each subscale the recidivism score(s) to which it relates and the number of underlying questionnaire items we can compute using our data. We use the data made available by ProPublica\footnote{
\href{https://github.com/propublica/compas-analysis}{https://github.com/propublica/compas-analysis}}; the Propublica dataset is missing features needed to compute the COMPAS score, and so we supplement this dataset with probation data from the Broward Clerk's Office. However, there remain missing items often related to subjective survey questions that cannot be computed without access to Northpointe's data, which are not publicly available. Notes on our data processing can be found in the supplement.

\begin{table}[htbp]
\caption{COMPAS subcales are inputs to the recidivism scores. We have some but not all of the questionnaire features that determine each subscale. For one of the History of Noncompliance features, ``Was this person on probation or parole at the time of the current offense?'', we can compute only whether the person was on probation.}
\label{table:subscales}
\centering
\begin{tabular}{lcc}
Subscale   & \# Features we have & Relevant \\ 
   &  / Total &  Recidivism Score \\ 
\\\hline
Criminal Involvement     & 4/4                         & General                   \\
History of Noncompliance & 3/5                         & Violent                   \\
History of Violence      & 8/9                         & Violent                   \\
Vocational/Educational   & 0/12                        & Both                      \\
Substance Abuse          & 0/10                        & General                  
\end{tabular}
\end{table}

According to our understanding of the COMPAS documentation, the general and violent COMPAS scores are both linear models, where age and age-at-first-arrest are the only factors that can have negative contributions to the COMPAS scores.

\subsection{COMPAS seems to depend nonlinearly on age, contradicting its documentation} \label{sec:howcompasdependsonage}

Let us consider if the linear model given by COMPAS is supported by our data. We make the following assumption:
\begin{quote}
\textbf{Data Assumption}: \textit{For most values of age, there is at least one person in our dataset with age-at-first-arrest equal to their age and the lowest possible risk score for each subscale.}
\end{quote}

First, note that age and age-at-first-arrest are the only COMPAS score inputs that have negative contribution to the COMPAS scores. Next, note that the least-risky value for age-at-first-arrest occurs when it is equal to current age (this implies that the individual has committed only one arrestable offense). Thus, individuals satisfying the Data Assumption should have the lowest COMPAS raw score for their age, which is key for the rest of our age analysis. In what follows, we describe attempts to confirm that these individuals are present in the data.

First, and most importantly, we used our data to check directly that for most values of age, there is at least one person in our dataset who has age-at-first-arrest equal to their age and who does not have criminal history. For the COMPAS general score, we can check if the Data Assumption holds for the Criminal Involvement subscale because we have all the inputs to it. We can only approximately check this assumption for the History of Violence and History of Noncompliance subscales because we do not have all the inputs. However, the Data Assumption seems plausible given that (1) the inputs to the subscales can take only a few integer values\footnote{In all but one case, the subscale inputs are binary or count variables that take up to only 6 values. The exception is the Criminal Involvement Subscale, which takes the total number of prior arrests, and for which we have data to directly validate.} and (2)  their typical content (e.g., ``Do you have a job?'' or ``How many times has this person been sentenced to jail for 30 days or more?'') suggests the least-risky input values are fairly likely. We cannot check the Data Assumption for the subscales for which we do not have data. However, the Data Assumption seems to hold for the subscales for which we \textit{do} have data (see Figure \ref{fig:check_dat_assump} in the appendix), leading us to believe it might hold generally.

\begin{figure}[htbp]
\centering
\begin{subfigure}{.5\columnwidth}
  \centering
  \includegraphics[width=1\linewidth]{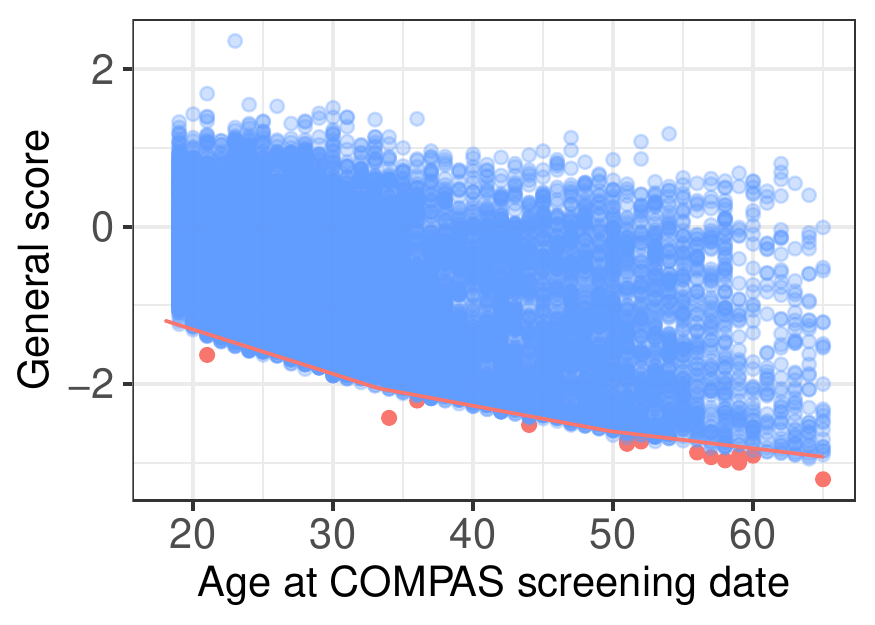}
  \caption{$f_{\textrm{age}}$}
  %\label{fig:sub11}
\end{subfigure}%
\begin{subfigure}{.5\columnwidth}
  \centering
  \includegraphics[width=1\columnwidth]{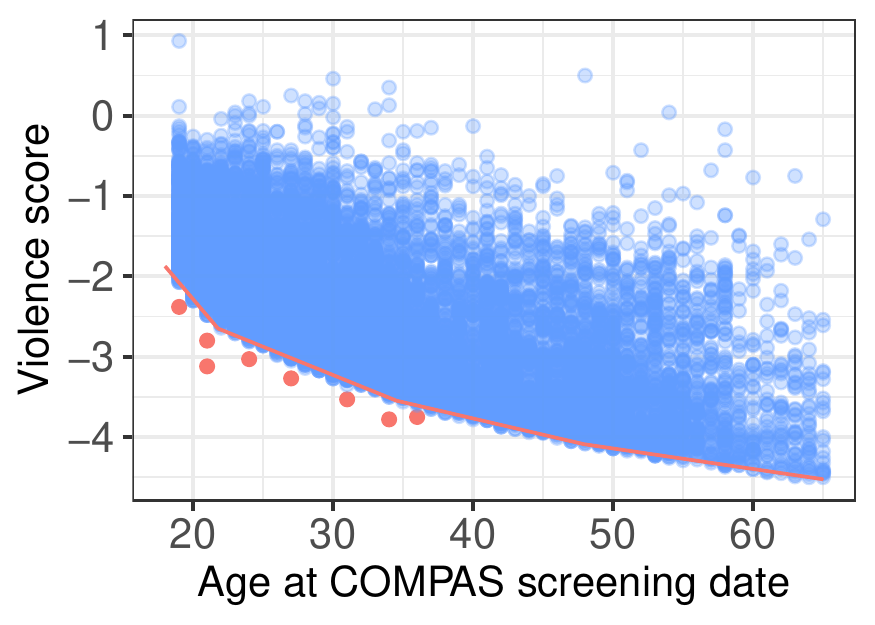}
  \caption{$f_{\textrm{viol age}}$}
  %\label{fig:sub22}
\end{subfigure}
\caption{\label{fig:agescatter} Scatter plot of COMPAS general recidivism score versus age and scatter plot of COMPAS violence recidivism score versus age. Age splines used to approximate the lower bound of the scatter plots are shown in red; age outliers that were removed from the analysis are also in red. }
\end{figure}

If the COMPAS documentation were correct in that COMPAS is a linear function of age, then, as long as the Data Assumption holds, if we plot the COMPAS raw score against age for all people, the lower bound should be a line with slope equal to the sum of the coefficients on age and age-at-first-arrest, since age equals age-at-first-arrest. Figure \ref{fig:agescatter} shows this is not the case. Also, the people near the lower bound of Figure \ref{fig:agescatter} often have no criminal history, and have age-at-first-arrest equal to age (see the appendix for more analysis). Thus, our evidence indicates the COMPAS model is not the linear model given above within the documentation. Recall that except for some typos, there should be no noise in the COMPAS scores. The COMPAS scores we observe should be the direct output of a computer program. 

Despite the lack of agreement of the data with the documentation, we would still like to reconstruct as much of the COMPAS black box as possible, so we make several weaker hypotheses about its form, none of which contradict the COMPAS documentation:

\begin{quote}
\textbf{Model Assumption 1}: \textit{The COMPAS raw score is an additive model with respect to each input (age, age-at-first-arrest, and the subscales).}

\textbf{Model Assumption 2}: \textit{The additive terms corresponding to each input except age and age-at-first-arrest are never negative.}

\textbf{Model Assumption 3}: \textit{For the lowest risk score, (a) the additive term corresponding to age-at-first-arrest is lowest when age-at-first-arrest is as large as possible (i.e., equal to age) and (b) the additive terms corresponding to the subscales are zero. }
\end{quote}

We believe Model Assumption 2 should hold because all of the inputs except age and age-at-first-arrest (e.g. number of violent felonies, number of times on probation) should lead to a higher risk of violence or recidivism, and therefore a higher COMPAS score.  Should Model Assumption 3 not hold, people with nonzero criminal history would have lower COMPAS scores than those with none, which is not intuitive. With these Model Assumptions and under the Data Assumption, the lower bound observed in Figure \ref{fig:agescatter} is exactly the additive term corresponding to age.

Reconstructing COMPAS's dependence on age is important because we know that COMPAS, if it is like other scoring systems, should depend heavily on age in order to predict well. If we could isolate and subtract off its dependence on age, we could more easily determine its dependence on protected attributes such as criminal history and race. Based on the lower bound of the COMPAS score with respect to age, we present a conjecture of approximately how the COMPAS score may depend on age, at least in Broward County, and we have a similar conjecture for the violence recidivism counterpart:\\

\noindent\textbf{\underline{{Conjecture:}}} \textit{The COMPAS general recidivism model is a nonlinear additive model. Its dependence on age in Broward County is approximately a linear spline, defined as follows:}
\begin{eqnarray*}
\textrm{for ages }\leq 33.26, f_{\age}(\age)=-0.056\times\age  -0.179\\
\textrm{for ages between 33.26 and 50.02},   f_{\age}(\age)=-0.032\times \age  -0.963\\
\textrm{for ages $\geq$ 50.02},   f_{\age}(\age)=-0.021\times\age  -1.541.
\end{eqnarray*}

%\begin{eqnarray*}
%\small{f_{\textrm{age}}(x)} &=& \small{  
%(4.4400685939 \times 10^{-4}) x^2
%- (7.2999449204 \times 10^{-2}) x 
%- 7.34079022286 \times 10^{-2}
%,}
%\end{eqnarray*}

%\end{eqnarray}
%\textit{where x is age. (Significant digits are kept because they are needed for the higher degree coefficients, and ensure replicability. Note that the polynomial is approximate, not exact due to the finite nature of data.) 
\textit{Similarly, the COMPAS violence recidivism model is a nonlinear additive model, with a dependence on age that is approximately a linear spline, defined by:}
%\begin{eqnarray*}
%\small{f_{\textrm{viol age}}(x)} &=& \small{ 
%(3.92860796271 \times 10^{-7}) x^4 
%- (8.27491652334 \times 10^{-5}) x^3} 
%\\ && \small{+ 
%(6.90196568436 \times 10^{-3}) x^2 
%- (2.96732870592 \times 10^{-1}) x
%+ 1.33552338832}.
%\end{eqnarray*}
\begin{eqnarray*}
\textrm{for ages }\leq 21.77, f_{\textrm{viol age}}(\age)= -0.205 \times\age + 1.815\\
\textrm{for ages between 21.77 and 34.58},   f_{\textrm{viol age}}(\age)= -0.070 \times \age  -1.113\\
\textrm{for ages between 34.58 and 48.36},   f_{\textrm{viol age}}(\age)=-0.040 \times \age  -2.166\\
\textrm{ for ages $\geq$ 48.36},   f_{\textrm{viol age}}(\age)= -0.025 \times \age -2.882.
\end{eqnarray*}

\begin{figure}[htbp]
\centering
\begin{subfigure}{.5\columnwidth}
  \centering
  \includegraphics[width=1\linewidth]{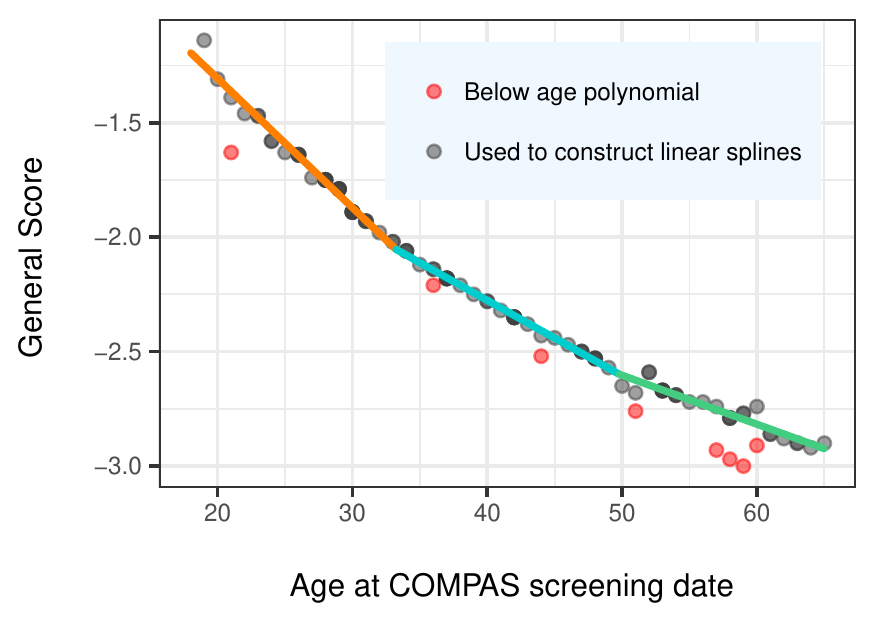}
  \caption{$f_{\textrm{age}}$}
  %\label{fig:sub11}
\end{subfigure}%
\begin{subfigure}{.5\columnwidth}
  \centering
  \includegraphics[width=1\columnwidth]{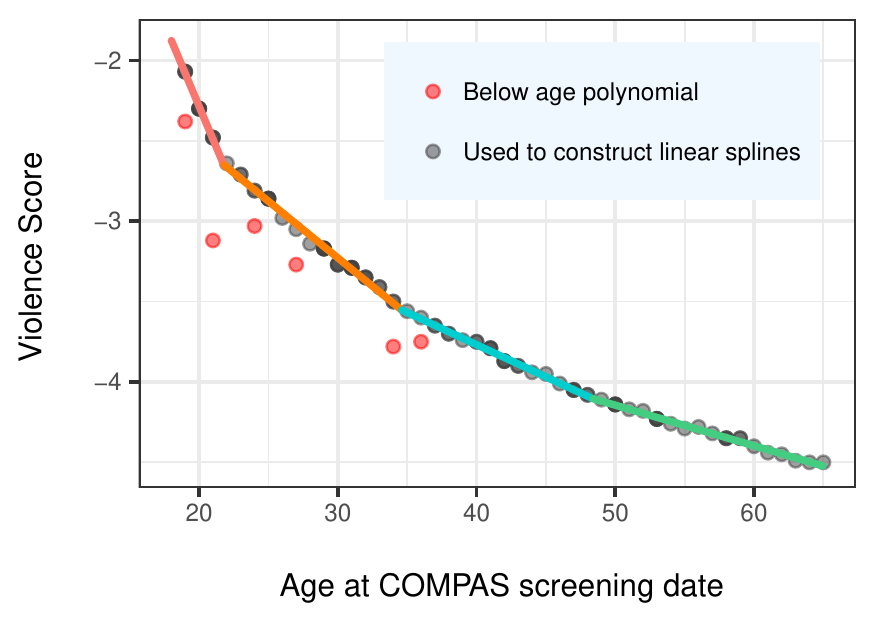}
  \caption{$f_{\textrm{viol age}}$}
  %\label{fig:sub22}
\end{subfigure}
\caption{Fitting
$f_{\textrm{age}}$ and $f_{\textrm{viol age}}$ to the data. The data are represented by the black dots and the fitting is shown as the multicolored line. Points which we deemed as age outliers are shown in red. 92 individuals were used to fit $f_{\textrm{viol age}}$ and 100 individuals were used to fit $f_{\textrm{age}}$. \label{fig:agesplines}}
\end{figure}

We used a two-stage procedure to fit the functions $f_{\textrm{age}}$ and $f_{\textrm{viol age}}$, with the goal of obtaining the closest approximation to the true functional relationship between age and the COMPAS raw scores as possible. First, we fit quadratic and quartic polynomials respectively to the lower bounds of the scatter plots of individuals' COMPAS general recidivism scores and the COMPAS violent recidivism scores. %, as depicted in Figure \ref{fig:agescatter}. 
Points more than $c = .05$ below these age polynomials were deemed ``age outliers" (a handful of individuals whose age seems to be incorrect) and removed from the analysis. Less than ten individuals (for each score) were removed due to this reason.
%If our model assumptions hold (age and age-at-first-arrest provide the only negative contributions to the COMPAS score) individuals with age equal to age-at-first-arrest and 0 for all other inputs to the COMPAS score should have the lowest possible score for their age. Thus, we filtered the data such that only the individuals who had zeros for the COMPAS inputs present in our data were included, and checked that most individuals on the lower bound of this filtered data have age equal to age-at-first-arrest. 
The age splines were fit using the lower bound of the subsets of individuals whom we hypothesize to satisfy the data assumption (individuals with age equal to age-at-first-arrest and no known contribution to any subscale), shown in Figure \ref{fig:agesplines}. In the left figure, several of the age outliers appear as if they were in a line, but if we add in the age outliers from the full population (not just those with no criminal history) the apparent line is no longer a lower bound.
 
 % \textcolor{red}{Generate new plots with age splines on the lb}
%The conjectures are shown using Figure \ref{fig:agesplines}; these are scatter plots of age versus the general recidivism COMPAS score for each individual in the ProPublica dataset on the left, and a similar plot on the right for the COMPAS violent recidivism score \textcolor{red}{The figure label is wrong, it is referring to Figure 1}. Functions $f_{\textrm{age}}$ and $f_{\textrm{viol age}}$ are shown as the colored piecewise linear lower bounds. Each individual, with very few exceptions, has a COMPAS general recidivism score that is at least as large as $f_{\textrm{age}}$, and a COMPAS violence score that is at least as large as $f_{\textrm{viol age}}$. 

% Figure \ref{fig:agesplines} shows the subset of data used to fit $f_{\textrm{viol age}}$. 

%We think it would be unlikely that the shape of the curves $f_{\textrm{age}}$ and $f_{\textrm{viol age}}$ are due to unmeasured variables. For the History of Violence subscale, we know from Northpointe documentation that all the components of the subscale take on values either 0, 1, 2, 3, 4, or 5 (some of them are binary, which means they are 0- or 1-valued). It would be unrealistic to assume that any of them would contribute negatively to the subscale score, based on their meaning -- they should all lead to a higher violence history score (e.g., it would be strange if ``number of past violent crimes" led to a lower score.) The individuals with the lowest scores for each age group tend to have low criminal histories (as shown in the supplementary materials), so any confounding variable producing this curve would need to smoothly vary with age, be somewhat unrelated to criminal history, and have a high weight in the COMPAS score in order to influence the shape of the curve. The existence of such as variable seems unlikely under these circumstances.

As discussed above, COMPAS' documentation claims a linear dependence on age. Even if COMPAS constructed the model to only take age as a linear argument, its predictions, as applied to a sample with representative covariate information, apparently induce nonlinear dependence on it. For the COMPAS documentation to claim such linear dependence can thus be misleading.

It is possible that age's \textit{only} contribution to the COMPAS general recidivism score is $f_{\textrm{age}}$ (similarly, $f_{\textrm{viol sage}}$ for the violence score). Let us describe why this seems to be true.
\noindent \textit{The remainders of general COMPAS minus $f_{\textrm{age}}$ and violent COMPAS minus $f_{\textrm{viol age}}$ do not seem to depend on age.}  After subtracting out the age polynomials, we employed  machine learning methods along with linear models (Table \ref{table:withandwithoutage}) to model the  remainder (COMPAS score minus the age polynomial). We ran each of these algorithms on the data, once using criminal history and age features only (\textit{with-age} models), and once using just criminal history (\textit{without-age} models). Machine learning methods are powerful, nonparametric models that are capable of modeling nonlinear functions very closely, given the right features. Thus, if the remainder depends on age coupled with criminal history, it is likely the accuracy will vary between the with-age and without-age models. However, instead, Tables \ref{table:withandwithoutage} and \ref{table:withandwithoutage_violent} (for the general and violence scores, respectively) show 
the accuracy of the machine learning models was almost identical between the with-age and without-age models. 

Importantly, if the dependence on age is additive, COMPAS does not use features that couple age and criminal history, such as the rate of crimes committed over time, despite the potential usefulness of this type of feature  (see, e.g., \cite{bushway2007inextricable}).
%There are arguments that coupling age with criminal history is useful to assess the rate at which crimes are committed; a younger person and an older person with the same number of past crimes are not committing crimes at equivalent rates \cite{bushway2007inextricable}.

\begin{table}[htbp]
\centering
\begin{tabular}{rc|c|c|c}
\multicolumn{1}{l}{}              
& Linear & Random  & Boosted  & SVM \\
& Model & Forest & Dec. Trees &  \\
\cline{2-5} 
\multicolumn{1}{r|}{Without Age} & 0.565        & 0.527         & 0.512 & 0.521  \\
\multicolumn{1}{r|}{With Age}    & 0.562        & 0.524         & 0.506 & 0.514
% The first row is Group 2, second row is Group 4 in code
\end{tabular}
\caption{Root mean square error for several machine learning algorithms for predicting COMPAS score minus age polynomial ($f_{\textrm{age}}$), with age included as a feature (bottom row), and without age (top row). We are trying to determine whether the COMPAS remainder (general COMPAS after subtracting the main age terms) still depends on age. The numbers for ``with age'' look very similar to the numbers ``without age.'' Thus, age does not seem to participate in the remainder term because accuracy does not change between the two rows. Race and age at first arrest are included as predictors for both ``with age'' and ``without age'' predictions.
\label{table:withandwithoutage}
}
\end{table}

\begin{table}[htbp]
\centering
\begin{tabular}{rc|c|c|c}
\multicolumn{1}{l}{}              
%& Linear Model & Random Forest & Boosted Decision Trees & SVM \\
& Linear & Random  & Boosted  & SVM \\
& Model & Forest & Dec. Trees &  \\
 \cline{2-5} 
\multicolumn{1}{r|}{Without Age} & 0.471        & 0.460         & 0.453 & 0.462  \\
\multicolumn{1}{r|}{With Age}    & 0.463        & 0.447         & 0.439 & 0.447
% The first row is Group 2, second row is Group 4 in code
\end{tabular}
\caption{Analogous to Table \ref{table:withandwithoutage} but for the COMPAS violence recidivism score, predicting COMPAS violence score minus $f_{\textrm{viol age}}$. Again, age does not seem to participate in this remainder.
\label{table:withandwithoutage_violent}
}
\end{table}

The fact that the lower bounds $f_{\textrm{age}}$ and $f_{\textrm{viol age}}$ seem to vary smoothly and uniformly with age, with only few outliers, indicates that the data entering into the COMPAS scores is high quality with respect to age. This has implications for our later analysis.

Now that we have hypothesized the dependence of COMPAS on age, we wish to explain its dependence on criminal history variables. We do this separately for the general and violence scores in Sections \ref{sec:crimhist_general} and \ref{sec:crimhist_violent}, respectively.

%%%%%%%%%%%%%%%
%\section{COMPAS has less dependence on race and criminal history than we expected}
%\label{sec:compascrimhist}

%After subtracting out the dominating age component $f_{\textrm{age}}$ from the COMPAS score and concluding that the remainder is unlikely to depend on age, we were hoping to iteratively reverse-engineer COMPAS, by subtracting out the remaining criminal history and other components in a similar manner. 
 
%According to the Northpointe documentation, the subscales comprising the rest of the score are linear with respect to their inputs. Since the criminal/probationary history data we have are inputs to these subscales, iteratively subtracting out these inputs from the COMPAS score remainders should succeed if the subscales are indeed linear. However, this approach was not successful. 

\subsection{Criminal history and the COMPAS general recidivism score}
\label{sec:crimhist_general}

\begin{figure}[htbp]
\centering
\begin{subfigure}{.48\columnwidth}
  \centering
  \includegraphics[width=1\columnwidth]{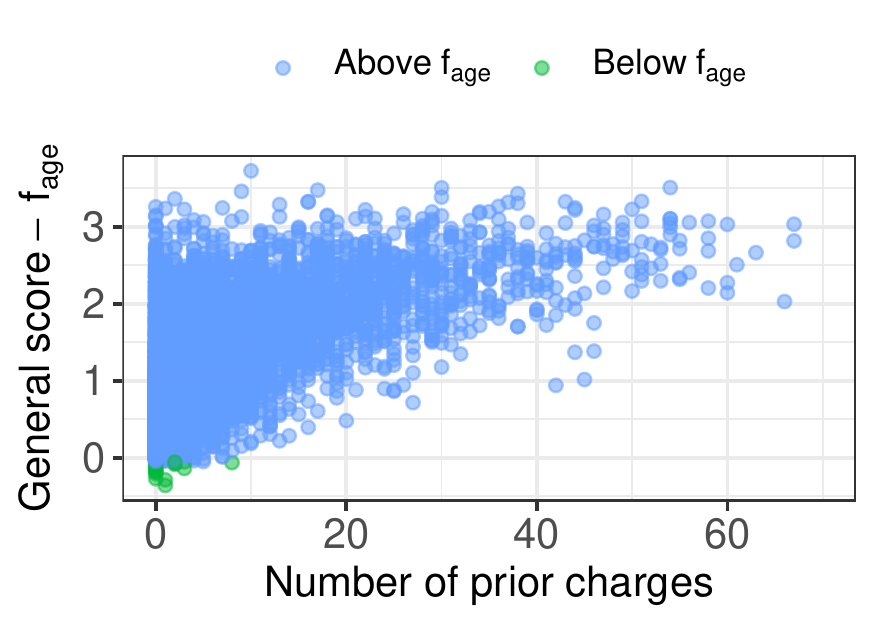}
  \caption{COMPAS $- f_{\textrm{age}}$ vs. number of priors. The green points are age outliers.}
  %\label{fig:sub1}
\end{subfigure} \hfill
\begin{subfigure}{.48\columnwidth}
  \centering
  \includegraphics[width=1\linewidth]{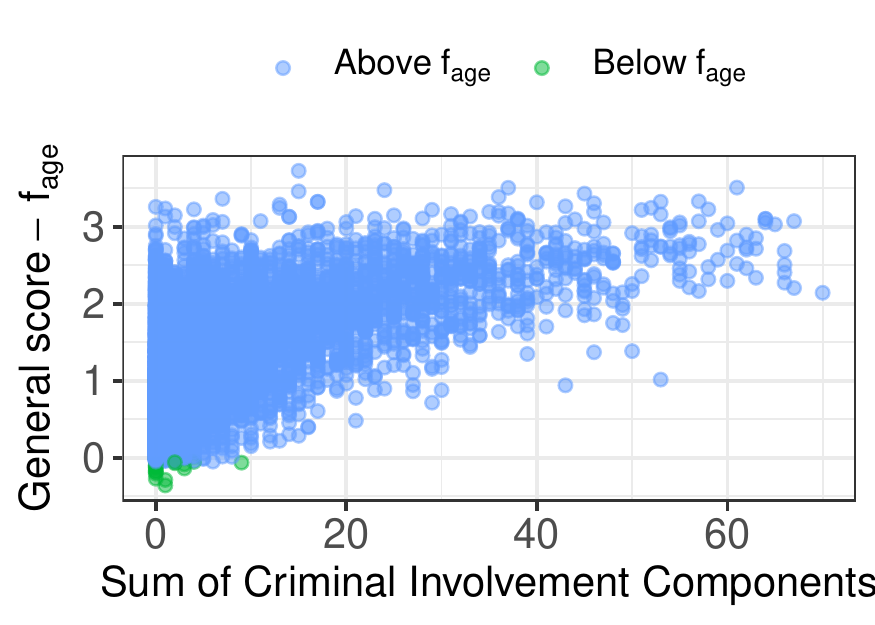}
  \caption{COMPAS $- f_{\textrm{age}}$ vs. unweighted sum of Criminal Involvement Subscale components. The green points are age outliers.}
  %\label{fig:sub2}
\end{subfigure}
\caption{We do not find a clear relationship between the COMPAS general score after subtracting the age spline and criminal history. Note that in each plot there are a few observations with a large number of prior charges that are outside of the plot range. 
	\textit{Left}: COMPAS $- f_{\textrm{age}}$ vs. number of priors. The green points are age outliers.
	\textit{Right}: COMPAS $- f_{\textrm{age}}$ vs. unweighted sum of Criminal Involvement Subscale components. 
 \label{fig:general_vs_crimhistory}}
\end{figure}

Unlike the dependence on age, the COMPAS general score does not seem to display a clear dependence on criminal history. Figure \ref{fig:general_vs_crimhistory} shows a scatter plot of COMPAS general score \textit{remainder} (which we define as the COMPAS score minus the age spline $f_{\textrm{age}}$) against the total number of prior charges, which is one of the variables determining the Criminal Involvement Subscale (left panel), and the unweighted sum of the variables in the Criminal Involvement Subscale (right panel). Note that we would ideally plot the remainder against the Criminal Involvement Subscale itself, but we do not know how the inputs are combined to form the subscale. Even excluding the age outliers (highlighted in green), there is no smooth lower bound as seen in Figure \ref{fig:agescatter}. Therefore we transition from searching for simple dependence of the COMPAS general score on its subscale items to searching for more complex dependencies.
%COMPAS general recidivism score does not seem to depend heavily on criminal history. Let us show this. Figure \ref{fig:modelremainder}(a) shows a scatter plot of number of prior charges versus the COMPAS general recidivism score remainder, COMPAS$-f_{\textrm{age}}$, where we colored the age outliers in red. Figure \ref{fig:modelremainder} (b) shows a scatter plot of general score remainder vs sum of criminal involvement items. 

\begin{figure}[htbp]
\centering
\begin{subfigure}{.5\columnwidth}
  \centering
  \includegraphics[width=.9\linewidth]{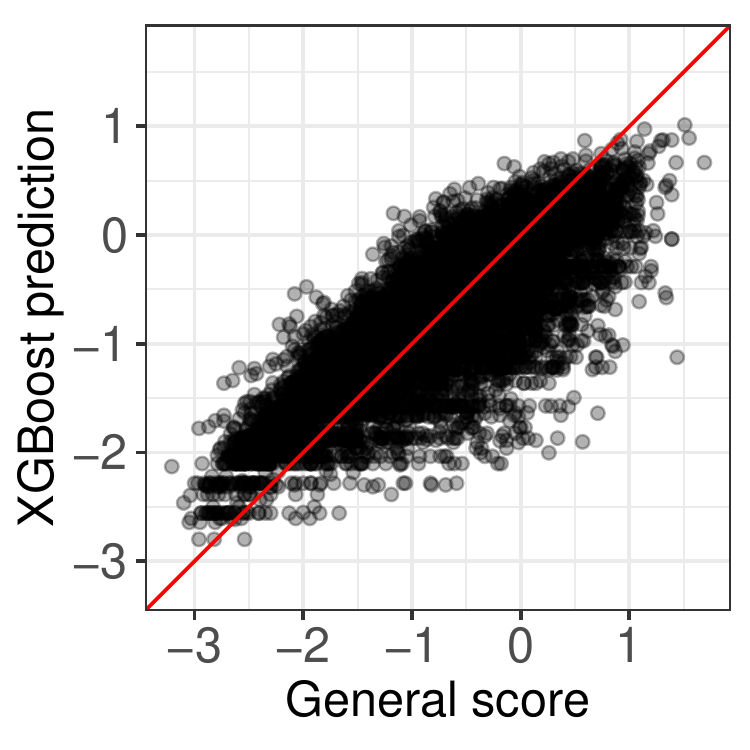}
  \caption{Predictions of COMPAS vs. actual values.}
  %\label{fig:sub1}
\end{subfigure}%
\begin{subfigure}{.5\columnwidth}
  \centering
  \includegraphics[width=.9\columnwidth]{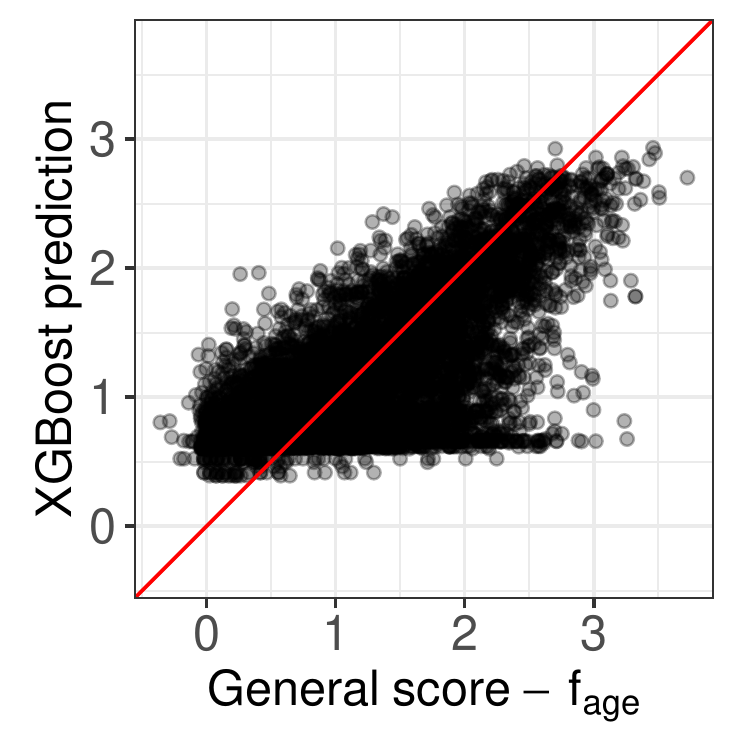}
  \caption{Predictions of COMPAS $-f_{\textrm{age}}$ vs. actual values.}
  %\label{fig:sub2}
\end{subfigure}
\caption{Predicting general remainder. \textit{Left}: Predictions of COMPAS vs. actual values.
\textit{Right}: Predictions of COMPAS $-f_{\textrm{age}}$ vs. actual values. \label{fig:modelremainder}}
\end{figure}

To then investigate whether the COMPAS general score depends in a more complex way on the Criminal Involvement Subscale items listed in the Appendix in Table \ref{table:features_crim}, we ran several machine learning algorithms (random forests \cite{randomForest}, boosted decision trees \cite{xgboost}, and support vector machines with a radial basis kernel function) on the subscale items our data has, to see if the COMPAS general recidivism score could be explained (either linearly or nonlinearly) by the subscale components. We tried predicting both the general score itself and the general score after subtracting $f_{\textrm{age}}$. Figure \ref{fig:modelremainder} shows a scatter plot of predictions versus the actual values for the two prediction problems. We make two observations from this figure. By comparing the two panels, we can see that the COMPAS general score seems to depend heavily on age, as the predictions of the COMPAS score remainder (right panel) are much worse than the predictions of the COMPAS score itself (left panel); this is because criminal history is correlated with age. After subtracting our reconstructed dependence on age (right panel), we see the ability of the criminal history variables to predict the COMPAS score remainder is surprisingly \textit{un}successful.
\textit{Thus the dependence of the COMPAS general score on criminal history, as captured by the components of the Criminal Involvement Subscale, seems to be weak.} 

% Notice that the first point above is different from the point of Table \ref{table:withandwithout}. Here we are saying the dependency on age is large. Previously, we showed it was correctly engineered. 

\subsection{Criminal history and the COMPAS violent recidivism score}
\label{sec:crimhist_violent}

\begin{figure}[htbp]
    \centering
    \includegraphics[width=.7\columnwidth]{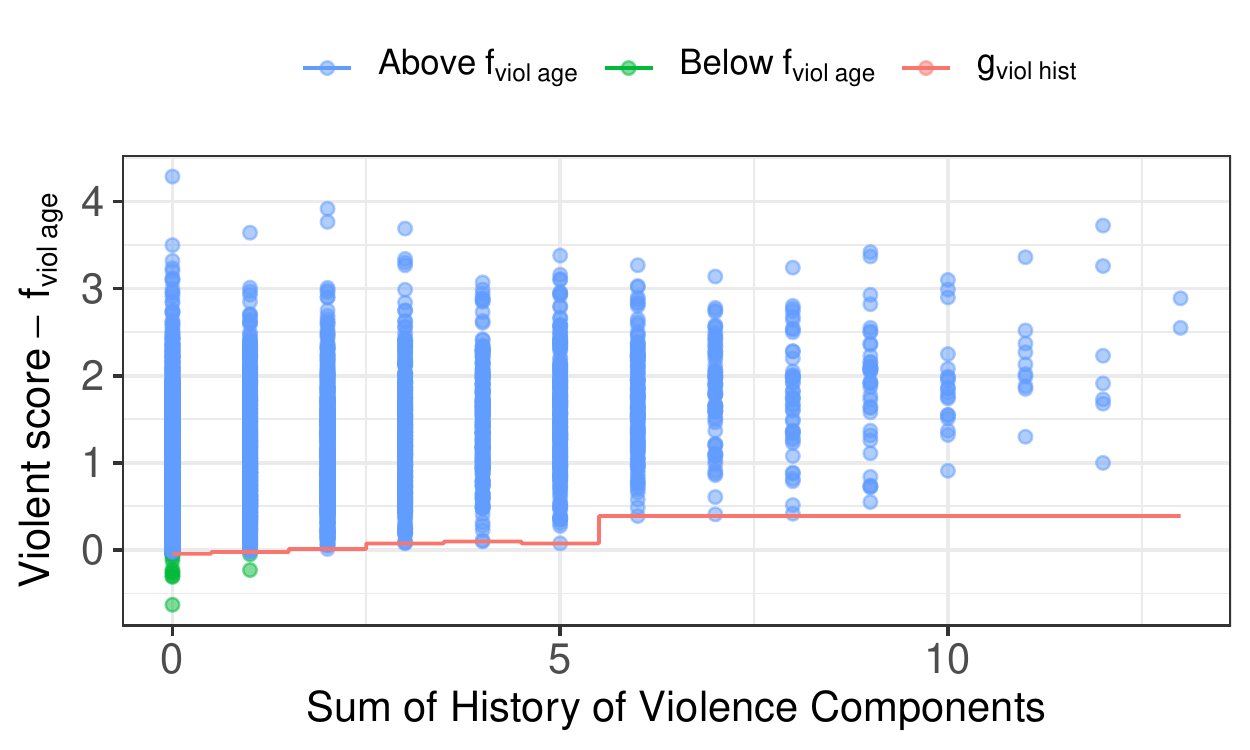}
    \caption{COMPAS $-f_{\textrm{age}}$ vs. sum of History of Violence components. Green points are age outliers.}
    \label{fig:violent_vs_viohist}
\end{figure}

\begin{figure}[htbp]
    \centering 
    \includegraphics[width=.7\columnwidth]{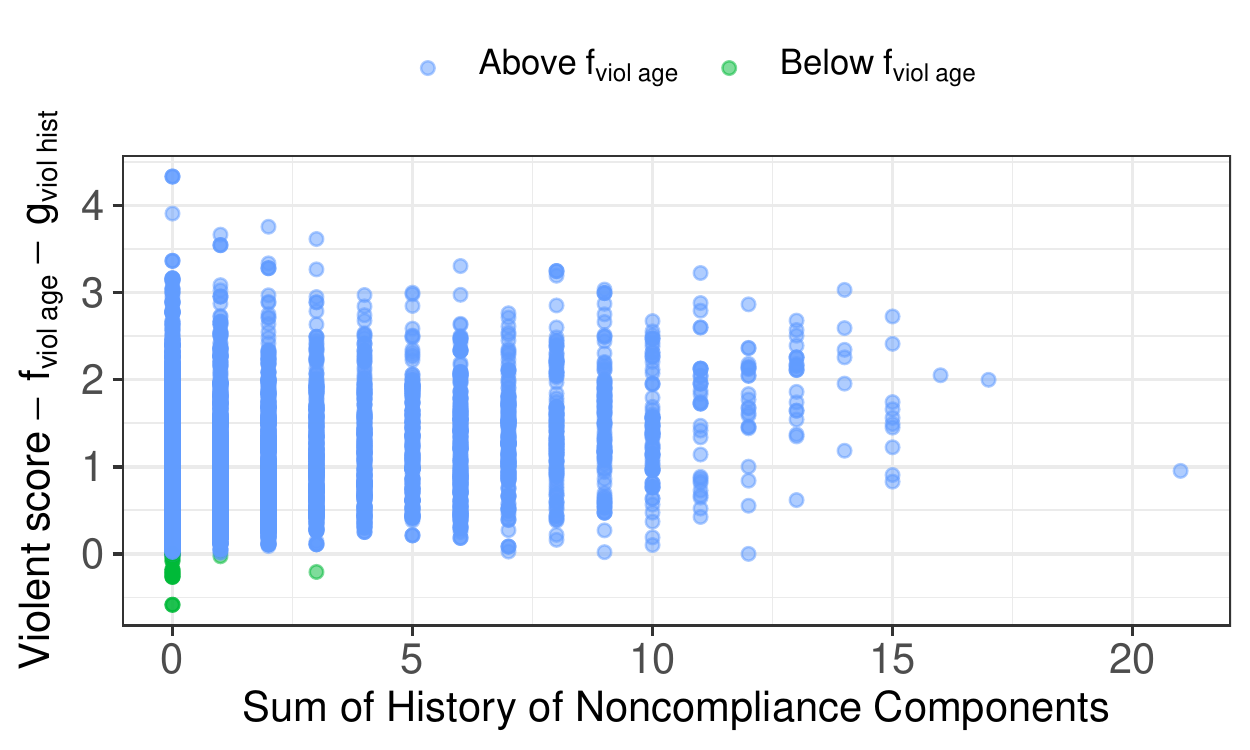}
        \caption{COMPAS $-f_{\textrm{age}} - g_{\textrm{viol hist}}$ vs. sum of History of Noncompliance components. Green points are age outliers.}
    \label{fig:violent_vs_noncomp}
\end{figure}

We gained more traction reconstructing the COMPAS violent recidivism score than the general score. Figure \ref{fig:violent_vs_viohist} shows the COMPAS violence score after subtracting the age spline $f_{\textrm{viol age}}$ against the unweighted sum of the Violence History Subscale components. 
Excluding the age outliers, this subtraction produced a crisp lower bound on the remainder, unlike the bounds we obtained trying various individual components and weighted sums of the components. We estimate the dependency on the Violence History Subscale as a piecewise linear function, which we call $g_{\textrm{viol hist}}$. Next, in Figure \ref{fig:violent_vs_noncomp}, we plot the remainder after also subtracting this dependency on Violence History (that is, the remainder of the COMPAS violence score after subtracting both $f_{\textrm{viol age}}$ and $g_{\textrm{viol hist}}$) against the unweighted sum of the components of the History of Noncompliance Subscale, on which the violence score should also depend. There is not a sharp lower bound that is consistent across the horizontal axis, which means this sum, by itself, is not likely to be an additive term within COMPAS. Therefore, we do not estimate a dependency on the unweighted sum of items in the History of Noncompliance Subscale. 

\begin{figure}[htbp]
\centering
\begin{subfigure}{.3\columnwidth}
  \centering
  \includegraphics[width=.9\columnwidth]{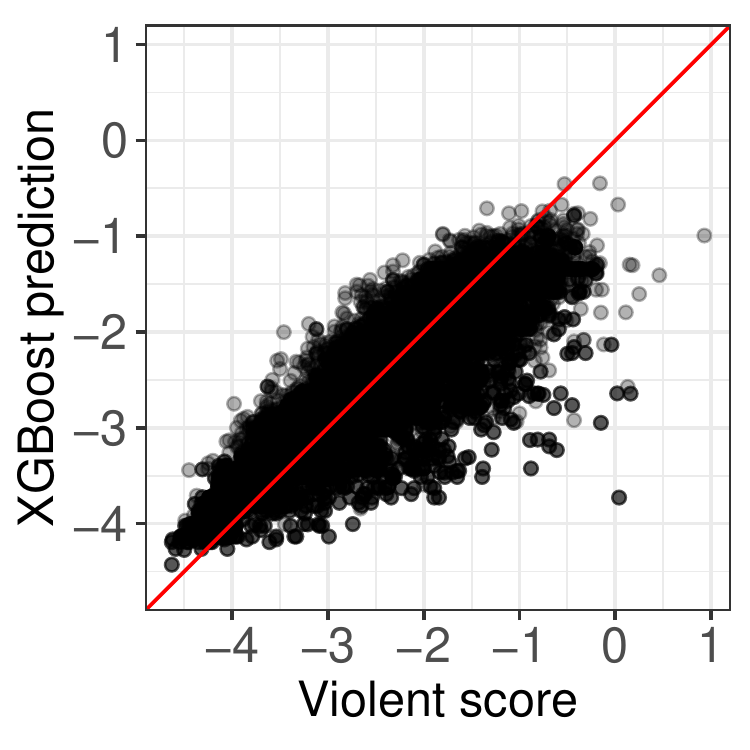}
  \caption{Predictions of COMPAS vs. actual values.}
  %\label{fig:sub1}
\end{subfigure} \hfill
\begin{subfigure}{.3\columnwidth}
  \centering
  \includegraphics[width=.9\columnwidth]{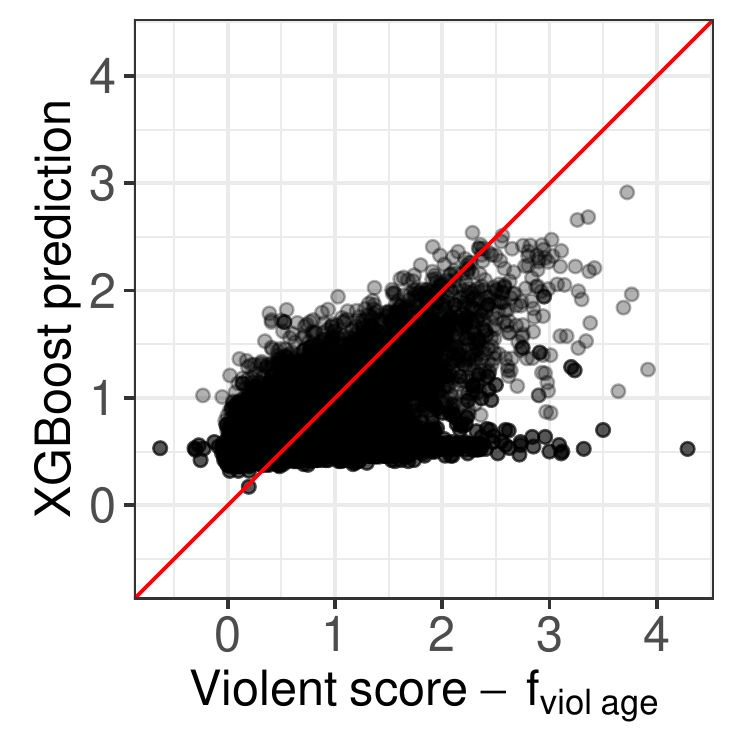}
  \caption{Predictions of $\text{COMPAS}-f_{\textrm{viol age}}$ vs. actual values.}
  \label{fig:sub2}
\end{subfigure} \hfill
\begin{subfigure}{.3\columnwidth}
  \centering
  \includegraphics[width=.9\columnwidth]{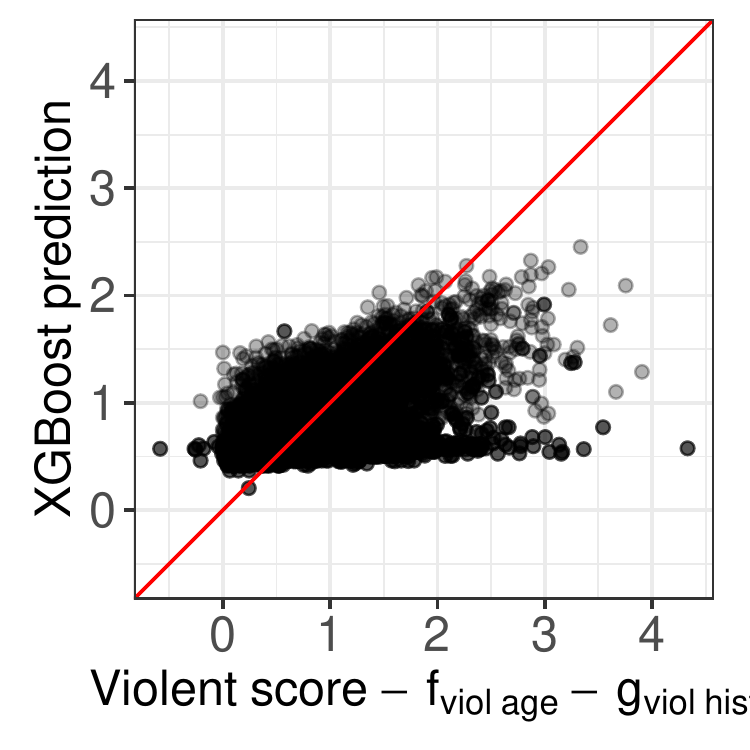}
  \caption{Predictions of $\text{COMPAS}-f_{\textrm{viol age}} - g_{\textrm{viol hist}}$ vs. actual values.}
  \label{fig:sub3}
\end{subfigure} \hfill
\caption{Predicting violent remainder. 
	\textit{Left}:   Predictions of COMPAS vs. actual values.
	\textit{Center}:   Predictions of $\text{COMPAS}-f_{\textrm{viol age}}$ vs. actual values.
	\textit{Right}:   Predictions of $\text{COMPAS}-f_{\textrm{viol age}} - g_{\textrm{viol hist}}$ vs. actual values.
 \label{fig:modelremainder_violent} }
\end{figure}

As with the COMPAS general score, we investigate if the COMPAS violence score depends on its subscale components in a more complex way. Figure \ref{fig:modelremainder_violent} shows the results of three separate prediction problems using all of the components in the History of Violence and History of Noncompliance subscales. From left to right, we use gradient boosted trees to predict the COMPAS violence score, the COMPAS violence score after subtracting $f_{\textrm{viol age}}$, and the COMPAS violence score after subtracting $f_{\textrm{viol age}}$ and $g_{\textrm{viol hist}}$. Comparing the panels in Figure \ref{fig:modelremainder_violent} from left to right, we see that the predictions degrade, emphasizing the importance of $f_{\textrm{viol age}}$ and $g_{\textrm{viol hist}}$, respectively, to the COMPAS violence score. That is, after subtracting these components, the input variables are much less able to predict the remaining COMPAS contribution. \textit{Thus, the dependence of the COMPAS violence score on criminal history, as captured by the Violence History and History of Noncompliance subscales, seems to be weak.}

\subsection{Caveats}
For both the general and violent COMPAS scores, we were unable to capture the remainder of the COMPAS score (i.e., after subtracting the reconstructed dependency on age) using the various criminal history variables that constitute the subscales. There could be several reasons for this, including the following, among other things:

\begin{itemize}
\item It is possible that our data are flawed. We obtained these data from a combination of ProPublica, the Broward County Sheriff, as well as the Broward County Clerk's office. We believe most of these data should be approximately the same as the data entered into the general COMPAS score. Furthermore, based on the analysis above, our age data on individuals seems to be high quality, so there is no \textit{a priori} reason that the criminal history data would be substantially lower quality. 

It is also possible that the way we calculated the criminal history subscale items for COMPAS differs from the way the Broward County Sheriff's Office calculates them. Our data processing is discussed in the appendix.

\item It is possible that we did not hypothesize the correct model form used in COMPAS; that is, our machine learning models may not have been able to express the nonlinearities present in COMPAS. While this could be true, we used very flexible models that should be able to fit (or even overfit) the COMPAS scores. Thus, we believe this is not a likely explanation.

\item It is possible that our data are incomplete. We know this is true, as COMPAS depends on factors other than criminal history. However, this leads to questions of what COMPAS can reasonably depend heavily on. Criminal history data are less noisy and less susceptible to manipulation than other survey questions; criminal history features do not depend on the survey-taker telling the truth about the answers. If COMPAS depended more heavily on survey questions than on criminal history, it could lead precisely to a kind of bias that we might want to avoid. For instance, if COMPAS did not depend heavily on the number of prior crimes, it might depend more heavily on socioeconomic proxies (e.g., ``How hard is it for you to find a job ABOVE minimum wage compared to others?" which is one of several questions on the COMPAS questionnaire that directly relates to socioeconomic status). 

\item There is something in the procedure of calculating  the COMPAS score that causes it to be calculated inconsistently. Since we do not know COMPAS, we cannot check this possibility. In the past, there have been documented cases where individuals have received incorrect COMPAS scores based on incorrect criminal history data \cite{nyt-computers-crim-justice,WexlerGlenn2017}, and no mechanism to correct it after a decision was made based on that incorrect score. We do not know whether this happens often enough to influence the scatter plot in a visible way. However, this type of miscalculation is one of the biggest dangers in the use of proprietary models. As we know from the calculations of other scoring systems besides COMPAS, if the number of crimes is not taken into account properly, or if the scoring system is calculated improperly in other ways, it could lead (and has led in the past) to unfair denial of parole, and dangerous criminals being released pre-trial.
%, and in one recent case, as discussed, this has led to a murder .\cite{npr-bail:2017, Ho2017}

\end{itemize}

Data quality issues, either for us or for the COMPAS score itself, are not just a problem for our analysis, but a problem that is likely to plague almost every jurisdiction that uses COMPAS or any other secret algorithm. This is discussed more in the next section.

%%%%%%%%%%%%%%%%%%%%
\subsection{Propublica seems to be incorrect in its claims of how COMPAS depends on race}\label{subsec:propublicarace} 

 If age and the components we have of the Criminal Involvement, History of Noncompliance, and History of Violence subscales can only explain COMPAS scores to a limited extent, then either the few components of these subscales that we are missing, or the remaining subscales, Substance Abuse for the violence score and Vocational/Educational for both scores, must be a large component of the scores. Reasoning these two subscales are highly correlated with race, we then tried to model the COMPAS remainders (i.e., after subtracting the age splines) with race as a feature, in addition to the available subscale components. Tables \ref{table:withandwithout} and \ref{table:withandwithout_violent}, respectively, show the results of several machine learning methods for predicting the general and violence score remainders. We see that these features cannot explain the COMPAS violence score remainders very well. \textit{Thus, to conclude, we hypothesize that COMPAS has at most weak dependence on race} after conditioning on age and criminal history.

\begin{table}
\centering
\begin{tabular}{rc|c|c|c}
\multicolumn{1}{l}{}              & Linear Model & Random Forest & Boosting & SVM \\ \cline{2-5} 
\multicolumn{1}{r|}{Without Race} & 0.572        & 0.533         & 0.520 & 0.522  \\
\multicolumn{1}{r|}{With Race}    & 0.562        & 0.524         & 0.506 & 0.514 
\end{tabular}
% First row is Group 3, second row is Group 4 in code.
\caption{RMSE of machine learning methods for predicting COMPAS general recidivism raw score after subtracting $f_{\textrm{age}}$ with and without race as a feature. There is little difference with and without race. The differences between algorithms are due to differences in model form. Age at COMPAS screening date and age-at-first-arrest are included as features.}
\label{table:withandwithout}
\end{table}

\begin{table}
\centering
\begin{tabular}{rc|c|c|c}
\multicolumn{1}{l}{}              & Linear Model & Random Forest & Boosting & SVM \\ \cline{2-5} 
\multicolumn{1}{r|}{Without Race} & 0.472        & 0.460         & 0.450 & 0.461  \\
\multicolumn{1}{r|}{With Race}    & 0.463        & 0.447         & 0.439 & 0.447 
\end{tabular}
% First row is Group 3, second row is Group 4 in code.
\caption{RMSE of machine learning methods for predicting COMPAS violence recidivism raw score after subtracting $f_{\textrm{viol age}}$ with and without race as a feature. Age at COMPAS screening date and age-at-first-arrest are included as features.}
\label{table:withandwithout_violent}
\end{table}

We replicated ProPublica's finding that a model with race predicts COMPAS well, but disagree with their conclusions. We repeated ProPublica's linear logistic regression on our slightly modified dataset, leading to a model, provided in the supplementary materials in Table \ref{table:propubregression}, whose race coefficient is large and significantly different from zero. Coefficients both for age and race are both large. 

There are several flaws in this analysis. First, the linearity assumption is likely to be wrong, as we know from considering the age analysis above. Second, the existence of an accurate model that depends on race is not sufficient to prove that COMPAS depends on race. Race is correlated with both criminal history and with age in this dataset. Because the linearity assumption is probably wrong, it is easily possible that race would appear to be significant, regardless of whether COMPAS is actually using race or its proxies (aside from criminal history and age) as important variables. As shown in Tables \ref{table:withandwithout} and \ref{table:withandwithout_violent}, including race as a variable to predict COMPAS does not improve prediction accuracy. That is, for each model we created that uses race, we found another almost equally accurate model that does not use race.
Thus, it is not clear that race or its proxies (aside from criminal history and age) are necessarily important factors in COMPAS.

In a separate analysis, Fisher et al$.$ \cite{FisherRuDo18} consider all models that approximate COMPAS with low loss, and among these, find models that depend the most and the least on race. 

%show that no accurate linear model for predicting general recidivism can have a high variable importance for the race variable, given age and criminal history. This result is not at odds with the ProPublica findings that we replicated. First,  Fisher et al$.$ \cite{FisherRuDo18}'s result is about predicting recidivism, whereas ProPublica was trying to predict the COMPAS score. Also, the statistical significance of the race variable in the ProPublica analysis is misleading because the linearity assumption underlying the regression does not hold. The result of \cite{FisherRuDo18} makes no such assumption, and does not compute significance. Their analysis relies only on permutation-based variable importance.

%%%%%%%%
\section{COMPAS sometimes labels individuals with long or serious criminal histories as low-risk}\label{sec:typos}

%As discussed earlier, one of the main uses of criminal scoring systems is to protect the public, so that individuals who pose a serious threat to society are not granted bail. 

We examine whether COMPAS scores can be low for individuals who pose a serious threat. Recently in California \cite{npr-bail:2017, Ho2017}, a defendant with a long criminal history was released pre-trial after a criminal history variable was inadvertently mistyped into a scoring system as being much lower than its true value. The defendant murdered a bystander before his trial. 

Typographical (data-entry) errors are extremely common \cite{Bushwayetal11}, which means risk-score errors occur regularly. For instance, if each person's COMPAS questionnaire contains 100+ questions, even a 1\% error rate could cause multiple wrong entries on almost every person's questionnaire. Data entry errors are a serious problem for medical records \cite{typos}, and in numerous other applications. The threat of typographical errors magnifies as the complexity of the scoring system increases; California had a very simple scoring system, and still typographical errors have caused serious events discussed above.

In what follows, we use real names and public records found easily on the Internet, following ProPublica, which is a news organization that compiled this public database and published real names and pictures of individuals within their report. It has been long debated whether this kind of information should be public, with citizen protection being a main argument (e.g., background checks for potential employees in sensitive jobs). There is an irony of this information being public, in contrast to the information actually used to make high-stakes decisions about these individuals being secret. However, this public information may allow us to determine that the secret information is potentially sometimes incorrect.

Consider the case of Desmond Rolle, whose criminal history included trafficking cocaine and aggravated battery of a pregnant woman, (felony battery  --- domestic battery by strangulation). He was given a COMPAS score of 1 (the lowest possible risk). A similar problem occurs for several individuals in the database. Table \ref{table:people} shows several such individuals who have COMPAS scores that appear to have been calculated with incorrect inputs, or whose criminal history information somehow has not been considered within the COMPAS formula. None of these individuals has scores below the age spline (none are age outliers).

%%% table was here!

While it is possible that COMPAS includes mitigating factors (employment, education, drug treatment) that reduce its score, it seems unlikely that they would reduce the score all the way to the lowest possible value, but since the model is not published, we cannot actually determine this. According to \cite{compas}, the only negatively weighted factors in COMPAS are age and age at first arrest, but according to our analysis above, these variables remain essentially constant with age for older individuals. This indicates there is no way to reduce a high score that might arise from a lengthy criminal history. Thus, what we are observing (long criminal histories with a low COMPAS violence score) should be impossible \textit{unless inputs have been entered incorrectly or omitted from the COMPAS score altogether}.

COMPAS general or violence scores do not include the current charges. Thus, in the case of Martin Owens in the ProPublica database, charged with a serious crime (kidnapping) but no prior crimes, he still receives the lowest-risk COMPAS score of 1. 

\begin{figure}[htbp]
\centering
\includegraphics[width=.9\columnwidth]{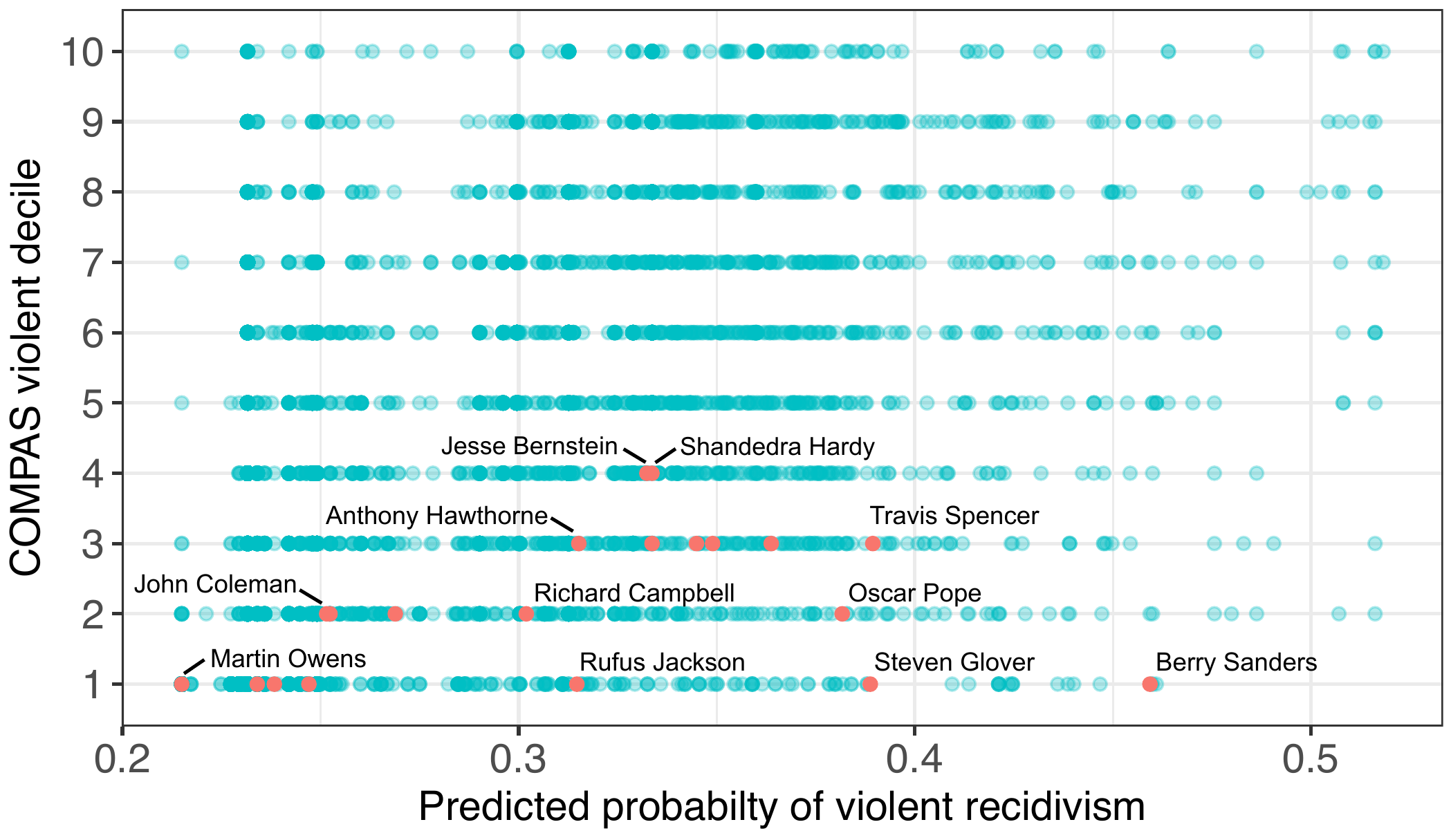}
\caption{Predicted probability of violent recidivism vs. COMPAS violence decile score. Individuals listed in Table \ref{table:people} are highlighted.\label{fig:longhist}}
\end{figure}

There are many individuals in the database whose COMPAS scores appear to be unreasonably high; however, it is possible that for those individuals, there are extra risk factors that cause them to be labeled high risk that are not in our database (e.g., incomplete criminal history information). Missing information would be able to explain COMPAS scores that seem too high, but it cannot explain COMPAS scores that are too low, such as the ones we presented above in Table \ref{table:people}. Figure \ref{fig:longhist} shows the predictions of a machine learning model versus COMPAS score. There are a huge number of individuals whose COMPAS score is much larger than the machine learning predictions, and also, there are many individuals for whom the machine learning model (a boosted decision tree) indicates high risk of recidivism, but the COMPAS score indicates a lower risk.

In cases like that of prisoner Glenn Rodr\"iguez whose parole was denied because of a miscalculated COMPAS score \cite{WexlerGlenn2017,nyt-computers-crim-justice}, he did not notice the error on his COMPAS form until after his parole was denied. 
Complicated forms, even if one is asked to check them over, lead to human error. We are certain, for instance, that there are still errors in this paper, no matter how many times we have checked it over -- and the longer the paper, the more errors it probably contains. 

Glenn Rodr\"iguez is a proven example of a data input error, but his case further demonstrates that \textit{data input transparency alone is insufficient to help wronged defendants}, the scoring mechanism must also be transparent. Mr. Rodr\"iguez was unable to change his parole sentence because he was unable to demonstrate that the data error impacted his COMPAS score.

\section{Is age unfair? Fairness through the lens of transparent models}\label{sec:ageonly}

Differences in true/false positive rates do not consider other factors like age and criminal history. In ProPublica's regression analysis of the COMPAS score, ProPublica conditioned on age and criminal history among other factors, indicating \textit{they thought COMPAS would hold to some notion of fairness had it depended only on age and criminal history}. (Otherwise, why condition on those two factors explicitly?). They used the significance of the coefficient on race to support their conclusion of bias against blacks. However, they used a linear term for age and did not handle unmeasured confounding, so this analysis was faulty. The faulty regression analysis leaves the differences in true/positive rates as their only remaining evidence of bias. However, the true/false positive rates are not conditioned on age and criminal history; that was why they performed the regression analysis, suggesting the regression could mitigate bias. In other words, Propublica's second analysis (the regression model) was invalid because it assumed a linear dependence on age. Their first analysis (the true/false positive rate analysis) would also then have been invalid for \textit{exactly the reasons why they conducted the second analysis} (which is that they would consider the model fair if COMPAS did not depend on race when conditioned on age and criminal history).

Age is a well-known determining risk factor for recidivism. Many recidivism scoring systems depend on age \cite{pasc2018valid,nafekh2002statistical,howard2009ogrs,Helmus12,langton2007actuarial,barnes2012classifying,hoffman1980salient,turner2009development,ZengUsRu2017} since it has no direct causal relationship with race (race does not cause age, age does not cause race), and it is a good predictor of future crime. For adults, the risk of recidivism decreases with age.\footnote{Figure \ref{fig:probrecid} in the appendix plots the probability of 2-year recidivism (defined by arrest within 2 years) as a function of age for individuals in Broward County, Florida, showing how it decreases as a function of age.}

On the other hand, in the Broward County data, African-American people tend to be disproportionately represented at younger ages than Caucasians; the median age of a COMPAS assessment on an African-American person is 27 years whereas the median age of a Caucasian is 33 years.\footnote{see the supplementary materials for full distributions}  
This means that more African-Americans will be labeled as high risk than Caucasians. This also means that more African-Americans will be \textit{mistakenly} labeled as high risk than caucasians. It also means that more Caucasians will be mistakenly labeled as low risk than African-Americans. Because of the difference in ages between blacks and whites in the dataset, even models that consider only age and criminal history can be as ``unfair'' as COMPAS by ProPublica's true/false positive rate analysis. 

\begin{figure}[htbp]
\centering
\includegraphics[width=.85\columnwidth]{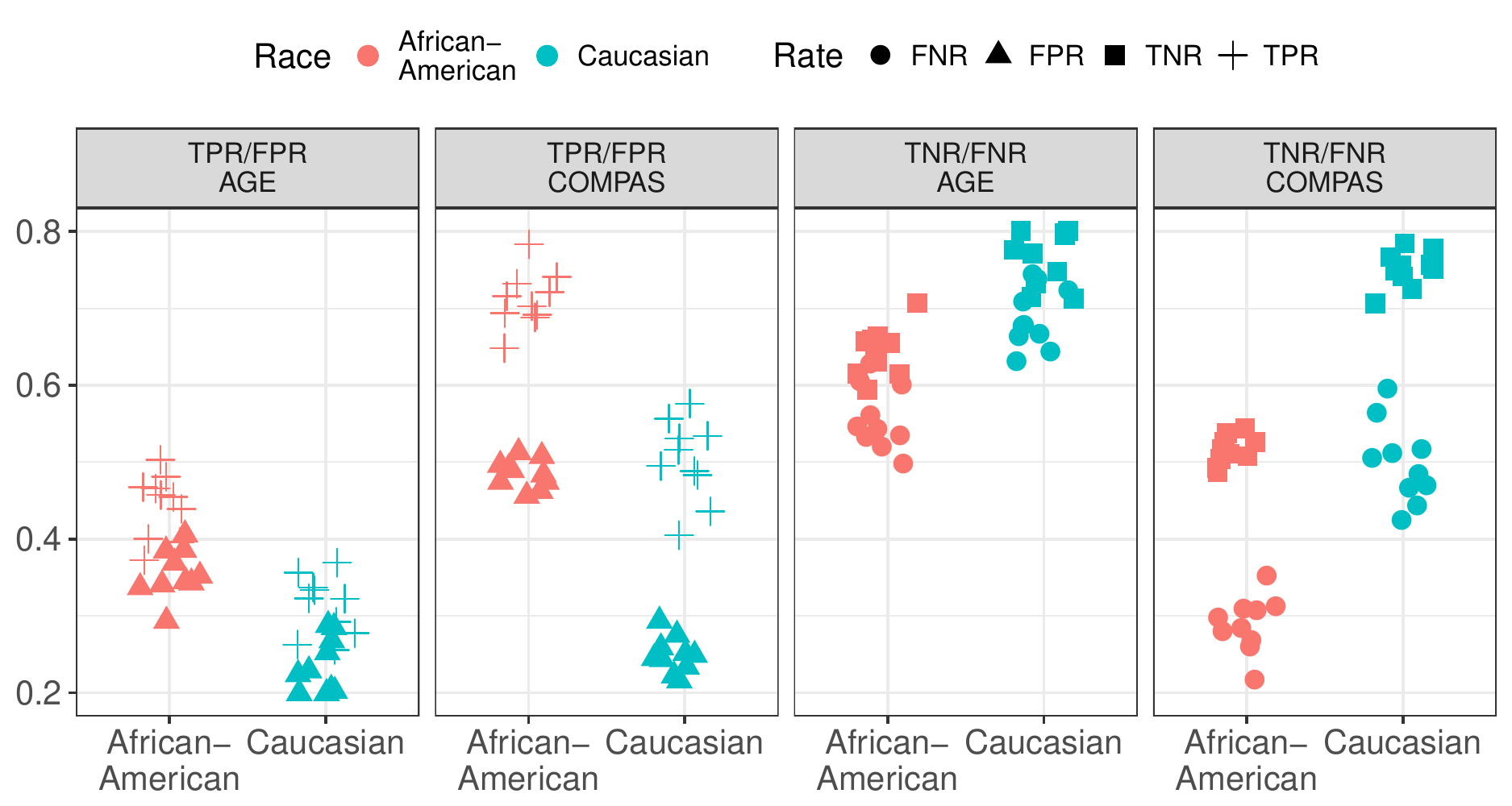}
\caption{Rates for the simple age model and for the COMPAS score. \textit{Age} appears also to be unfair.\label{fig:tpr}}
\end{figure}

Figure \ref{fig:tpr} shows the true positive rate (TPR), false positive rate (FPR), true negative rate (TNR) and false negative rates (FNR) for the model \textit{age}, which is defined to be ``If age $\leq$ 24, then predict arrest within 2 years, otherwise predict no arrest." 
The figure also shows the rates for the COMPAS general recidivism score. The data were divided into 10 randomly chosen folds, and the rates are plotted for all folds, showing a consistent pattern across folds. Indeed, we observe higher false positive rates for African-Americans, and higher false negative rates for Caucasians. There is an elevated $\approx$10\% higher FPR for African-Americans than for Caucasians for \textit{age}, and a $\approx$10\% higher FNR for Caucasians than African-Americans for age. These differing levels constitute one of the accusations of unfairness by ProPublica, which means that \textit{age} is an unfair risk prediction model by this definition. COMPAS seems to be more ``unfair" than \textit{age}, but as we have seen, it may be possible to explain this unfairness by a combination of age and other features that differ between the distributions of African-Americans and Caucasians and have little to do with the COMPAS score itself. In fact, we also know from \cite{angelino2018} that a very simple model involving age and the number of priors is just as unfair as COMPAS by this definition. 

The many definitions of fairness often contradict with each other. Had ProPublica considered just the transparent model \textit{age} (instead of analyzing COMPAS), it would have been easy to see that the difference in age populations between African-Americans and Caucasians caused their two notions of fairness to disagree. In that case, would they still have written an article claiming that the difference in true and false positive rates meant that COMPAS was unfair? Or would they have claimed that the use of age was unfair rather than conditioning on it? Or would they have claimed the data collection process was unfair? Using a transparent model can sometimes enlighten an understanding of the debate on fairness.

%To summarize this argument: transparency allowed us to determine that ProPublica's two fairness analyses seem to be at odds. By controlling for age and criminal history, and then looking for dependence on race, it indicates that the use of age could be considered fair. the knowledge that (1) COMPAS depends heavily on age, (2) age itself creates ``unfair" models

The consequences of using age in criminal risk assessments are explained nicely by \cite{Patrick2018}, however, the use of criminal history to assess fairness is confusing for additional reasons. If we do use criminal history in risk prediction, since African-Americans tend to have longer criminal histories, their scores will be worse. On the other hand,  
%This logic carries over to criminal history as well. People with long criminal histories are more likely to commit further crime in the future. By ProPublica's definition of fairness, using criminal history is unfair because criminal history correlates with race. However, 
if we do not use criminal history, our risk predictions would be worse. In that case, we could be releasing dangerous criminals based on poor pre-trial risk assessments, which leads to poor decisions for the public. 
 %However, if age is not considered fair, and if one insists on using ProPublica's definition of fairness, then there would not exist a model that is both accurate and fair. 

Of course, if we eliminate the most important predictors of recidivism (age and criminal history) on grounds of leading to unfair models, it is not clear that any useful predictors of criminal recidivism remain. In that case, we could be stuck in the situation where a human decision-maker provides non-transparent, potentially biased decisions.

The point of this exercise is not to determine whether the age model is fair by any given definition -- the age model is transparent, which makes it much easier to debate, and useful for explaining different possible definitions of fairness and how they may never intersect. ProPublica's regression analysis seems to assume that using age in a risk prediction model is reasonable. But \textit{is} age unfair? If we cannot decide on whether the age model is fair, we certainly cannot decide on whether COMPAS is unfair. However, it is certainly much easier to debate about the transparent and simple age model than about a black-box scoring system. While a review of the numerous definitions of fairness \cite{FATML,berk2017} is outside the scope of this work, a potentially easy way to alter the definition of fairness is to control for non-protected covariates such as age. In that case, as long as predictions for African-Americans and Caucasians have equal true/false positive rates for each age group, then the model would be considered fair. Of course, a problem with this definition is that any policy that targets young people disproportionately affects African-Americans.

%Let us assume that using age in a model is considered to be fair. In that case, we cannot use the definition of fairness used by ProPublica (even though it is common) because age correlates with race in the criminal population. If we did use Propublica's definition, it would be incompatible with the age model being fair. In that case, no recidivism model used currently would be ``fair" --- all use age. 

% Transparency as a type of procedural fairness

\section{Discussion} % Lessons learned}

After attempting to isolate COMPAS' dependence on age, we were able to investigate how much COMPAS can depend on criminal history and proxies for race. We found that it is unlikely that COMPAS depends heavily on either of them. Machine learning methods for predicting COMPAS scores performed equally well with or without direct knowledge of race. This seems to contradict ProPublica's claims, but ProPublica's methodological assumptions (at least about COMPAS depending linearly with age) were wrong, which caused their conclusions to be faulty. 

Northpointe claims the current charge is not helpful for prediction of future violent offenses \cite{compas}. (Oddly, they have a separate ``Current Violence" scale that includes the current charges, but which is not claimed to be predictive.) How much should one weigh the current charges with the COMPAS scores? This is not clear. Because COMPAS is a black box, it is difficult for practitioners to combine the current charge (or any other outside information) with the COMPAS scores. Because the current charges are separate, COMPAS scores are not single numbers that represents risk. Instead their interpretation has a large degree of freedom. Could  decision-makers fail to realize that the COMPAS score does not include the current charge? Perhaps this alone could lead to faulty decision-making.

We showed examples where COMPAS scores can label individuals with long criminal histories as low-risk. This could easily stem from a lack of transparency in COMPAS and could lead to dangerous situations for the public. Even if COMPAS were completely fair, by some reasonable definition of fairness, this would not stop it from being miscalculated, leading to a form of procedural unfairness. Since it is known that COMPAS is no more useful for predicting recidivism than simple, interpretable models, there is no good reason to continue using complicated, expensive, error-prone proprietary models for this purpose. There is a mystique behind the use of black box models for prediction. However, just because a model is a proprietary does not mean it is any better than a publicly available model \cite{Rudin18}.

Interestingly, a system that relies only on judges -- and does not use machine learning at all -- has similar disadvantages to COMPAS; the thought processes of judges is (like COMPAS) a black box that provides inconsistent error-prone decisions. Removing COMPAS from the criminal justice system, without a transparent alternative, would still leave us with a black box.

Privacy of data should be more of a concern than it presently is. If COMPAS does not depend heavily on most of the 137 variables, including the proxies for socioeconomic status, it is not clear if Northpointe is justified in collecting such private information. COMPAS is a risk \textit{and needs} assessment, but is that private information necessary to assess an individual's needs? All evidence suggests it does not seem to be necessary for estimating risk. Determination of needs seems to be a complicated causal question about who benefits from what types of treatments.  This issue is beyond the scope of this article, but is important. Northpointe's control over criminal risk scores is analogous to Equifax's control over credit scores, and leads to inherent privacy risks.

Thus, our findings indicate that some form of unfairness caused by COMPAS can affect almost everyone involved in the justice system: 1) lack of transparency makes it difficult to assess any of the myriad forms of fairness, leading to faulty arguments like those of ProPublica. \textit{Lack of transparency can hide bias towards underrepresented groups, or conversely, it can make fair models seem biased.} 2) The unnecessary complexity of COMPAS could cause injustice to those who had typos in the COMPAS computation, and as a result, were given extra long sentences or denial of parole. (\textit{This is procedural unfairness}.) 3) It is possibly unfair to the taxpayers and judicial system to pay for the collection of long COMPAS surveys and COMPAS predictions when simpler, transparent, options are available. (\textit{Poor designation of public resources is unfair to everyone}.) 4) The subgroup of people who provided very private personal information (e.g., about their family history of crime or poverty) to Northpointe has potentially been wronged. (\textit{There is a form of privacy unfairness in being forced to provide personal information to an entity when it is unnecessary to do so.})

The problems with COMPAS pertain to many industries. Without community standards or policy requirements for transparency, business considerations disincentivize creators of models to disclose their formulas. However, this lack of transparency is precisely what allows error to propagate and results in damage to society \cite{Rudin18}. Merely being able to explain black box models is not sufficient to resolve this -- the models need to be fully transparent, and in criminal justice, there is no loss in predictive accuracy for using a transparent model.

%The ProPublica article about COMPAS has led to a heated discussion about fairness, when there are much more serious issues due to lack of transparency; fairness is much more easily discussed when models are transparent, and typos due to lack of transparency are unacceptable. COMPAS seems to depend highly on age, and possibly less on criminal history and race than was claimed by ProPublica.

%This analysis indicates firstly, that the definition of fairness used by ProPublica and others is problematic. This definition also states that \textit{age} alone is an unfair predictor of recidivism. If one believes that using age alone leads to fair models, then there is a problem with the ProPublica definition of fairness. If one instead agrees with ProPublica's fairness definition, then since criminal history has the same characteristics as age, \textit{no model} using age or criminal history is allowed, in which case there is little point trying to construct a risk prediction model at all, as it will either be inaccurate, or unfair.

\subsection*{Code}
Our code is here: \url{https://github.com/beauCoker/age_of_unfairness}

%% References
% Your references go at the end of the main text, and before the
% figures.  For this document we've used BibTeX, the .bib file
% scibib.bib, and the .bst file Science.bst.  The package scicite.sty
% was included to format the reference numbers according to *Science*
% style.

%BibTeX users: After compilation, comment out the following two lines and paste in
% the generated .bbl file. 
%%% Use the first one for KDD version
%\bibliographystyle{ACM-Reference-Format}
\bibliographystyle{Science_v2}
\bibliography{recidivism}

%\textcolor{red}{This is where Table 6 is - need to fix}
\input{tableindividuals}

\onecolumn
\section{Acknowledgments}
We acknowledge partial funding from the Laura and John Arnold Foundation. Thank you to the Broward County Sheriff's office.

\section*{Supplementary materials}

\subsection*{Supporting figures and tables}

\subsubsection*{Probability of recidivism as a function of age}

\begin{figure}[htbp]
\centering
\includegraphics[width=0.5\textwidth]{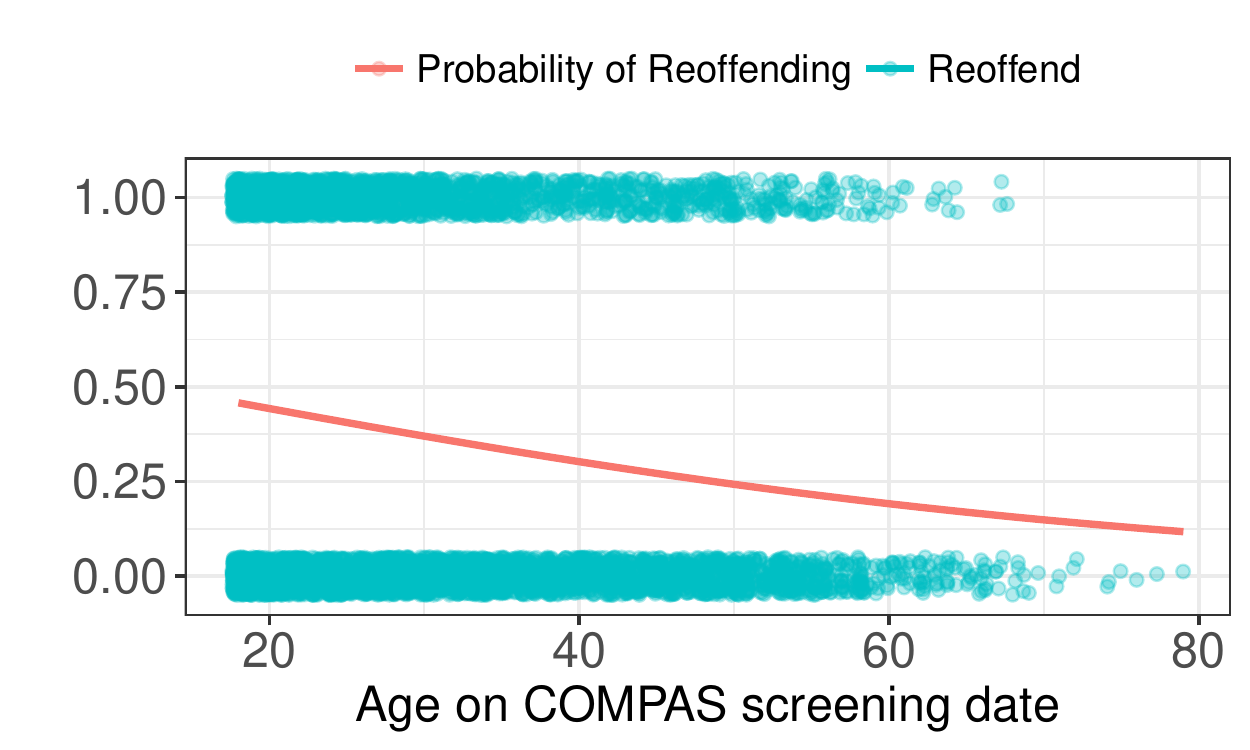}
\caption{Probability of charge within 2 years as a function of age (red). The blue scatter plot is useful for understanding the distribution of ages of individuals; each individual who was arrested has a dot at their age on the horizontal axis, and a ``1" on the vertical axis.\label{fig:probrecid}}
\end{figure}

Figure \ref{fig:probrecid} shows the probability of a new charge within 2 years as a function of age. The probability is a decreasing function of age.

\subsubsection*{Age histograms}

\begin{figure}[htbp]
\centering
\includegraphics[width=.7\textwidth]{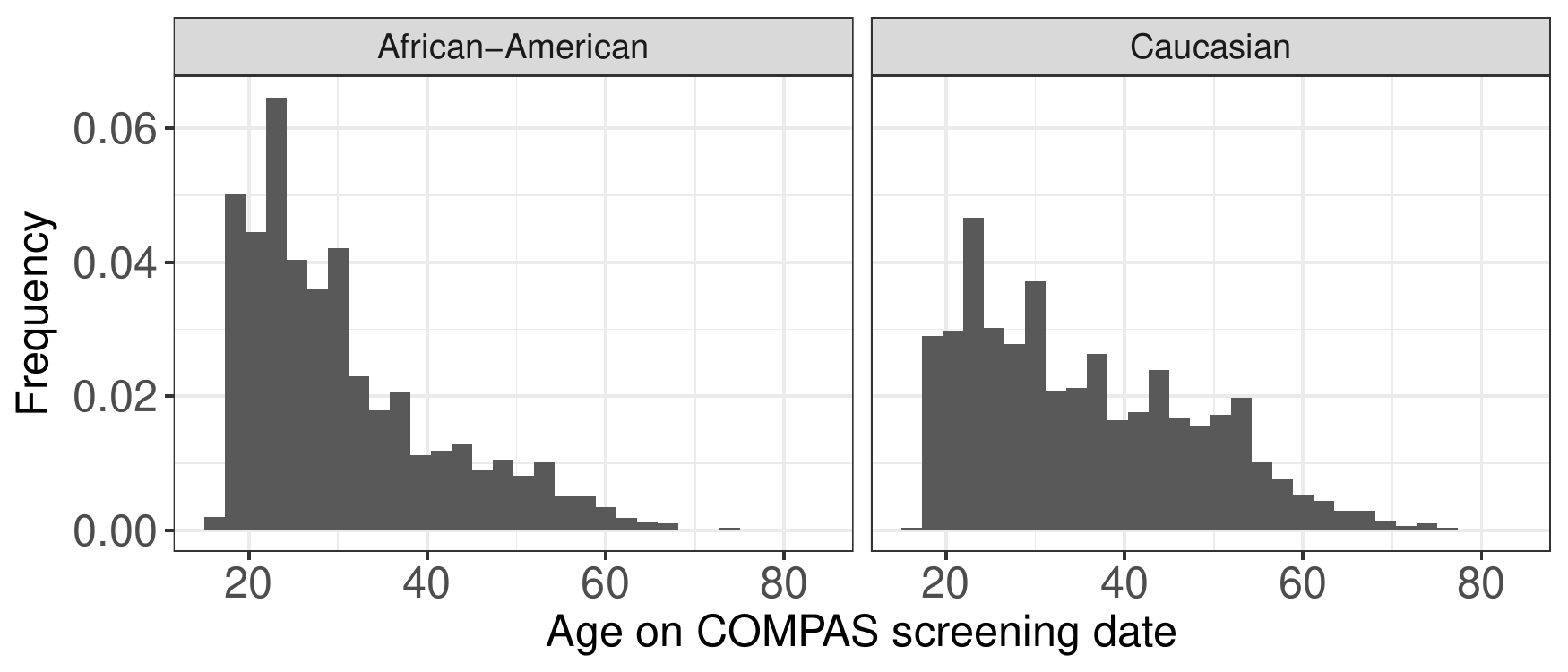}
\caption{Normalized histograms of age for African-Americans and Caucasians within the Broward County dataset.\label{fig:histo}}
\end{figure}

Figure \ref{fig:histo} shows the normalized histograms of African-Americans and Caucasians within Broward County who were evaluated with COMPAS between the beginning of 2013 and the end of 2014. These histograms do not involve COMPAS scores themselves, only information about the set of individuals who received COMPAS scores. The histogram for African-Americans is skewed to the left, which means African-Americans tend to be younger on average when their COMPAS score is calculated in the Broward County dataset. 

%%%%%%%%%%%%%%%%%%%%%%

\subsubsection*{Predictions of recidivism with and without race}

Tables \ref{table:recid_withandwithout} and \ref{table:recid_withandwithout_violent} show predictions of recidivism and violent recidivism, respectively, with and without race as a feature. The results are very similar with and without race. 

\begin{table}[h!]
\centering
\begin{tabular}{rc|c|c|c}
\multicolumn{1}{l}{}              & Logistic Regression & Random Forest & Boosting & SVM \\ \cline{2-5} 
\multicolumn{1}{r|}{Without Race} & 0.330        & 0.321         & 0.312 & 0.298  \\
\multicolumn{1}{r|}{With Race}    & 0.328        & 0.318         & 0.309 & 0.313 
\end{tabular}
% First row is Group 3, second row is Group 4 in code.
\caption{Misclassification error of machine learning methods for predicting general recidivism with and without race as a feature. Age at COMPAS screening date and age-at-first-arrest are included as features. Unlike when predicting the COMPAS raw score remainder, we include the current offense in criminal history features.}
\label{table:recid_withandwithout}
\end{table}

\begin{table}[h!]
\centering
\begin{tabular}{rc|c|c|c}
\multicolumn{1}{l}{}              & Logistic Regression & Random Forest & Boosting & SVM \\ \cline{2-5} 
\multicolumn{1}{r|}{Without Race} & 0.159        & 0.165         & 0.158 & 0.160  \\
\multicolumn{1}{r|}{With Race}    & 0.159        & 0.161         & 0.160 & 0.160 
\end{tabular}
% First row is Group 3, second row is Group 4 in code.
\caption{Misclassification error of machine learning methods for predicting violent recidivism with and without race as a feature. Age at COMPAS screening date and age-at-first-arrest are included as features. Unlike when predicting the COMPAS raw score remainder, we include the current offense in criminal history features.}
\label{table:recid_withandwithout_violent}
\end{table}

%%%%%%%%%%%%%%%%%%%%%%%%%%%
\subsection*{Fitting $f_{\textrm{age}}$ and $ f_{\textrm{viol age}}$}

%%%%%%%%%%%%%%%%%
\subsubsection*{Checking Data Assumptions}

\begin{figure}[htbp]
\centering
\begin{subfigure}{.48\columnwidth}
  \centering
  \includegraphics[width=1\columnwidth]{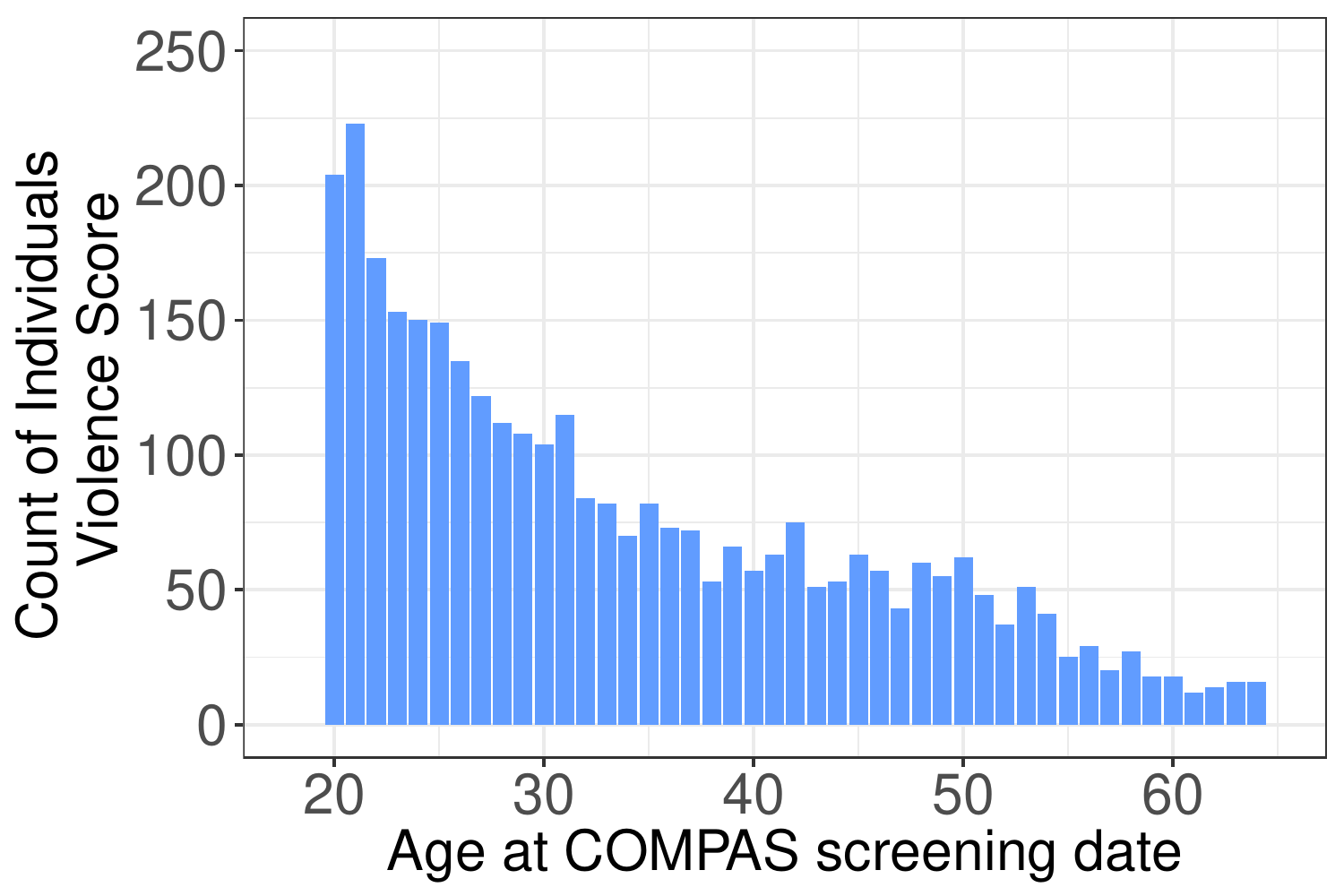}
  \caption{}
\end{subfigure} \hfill
\begin{subfigure}{.48\columnwidth}
  \centering
  \includegraphics[width=1\linewidth]{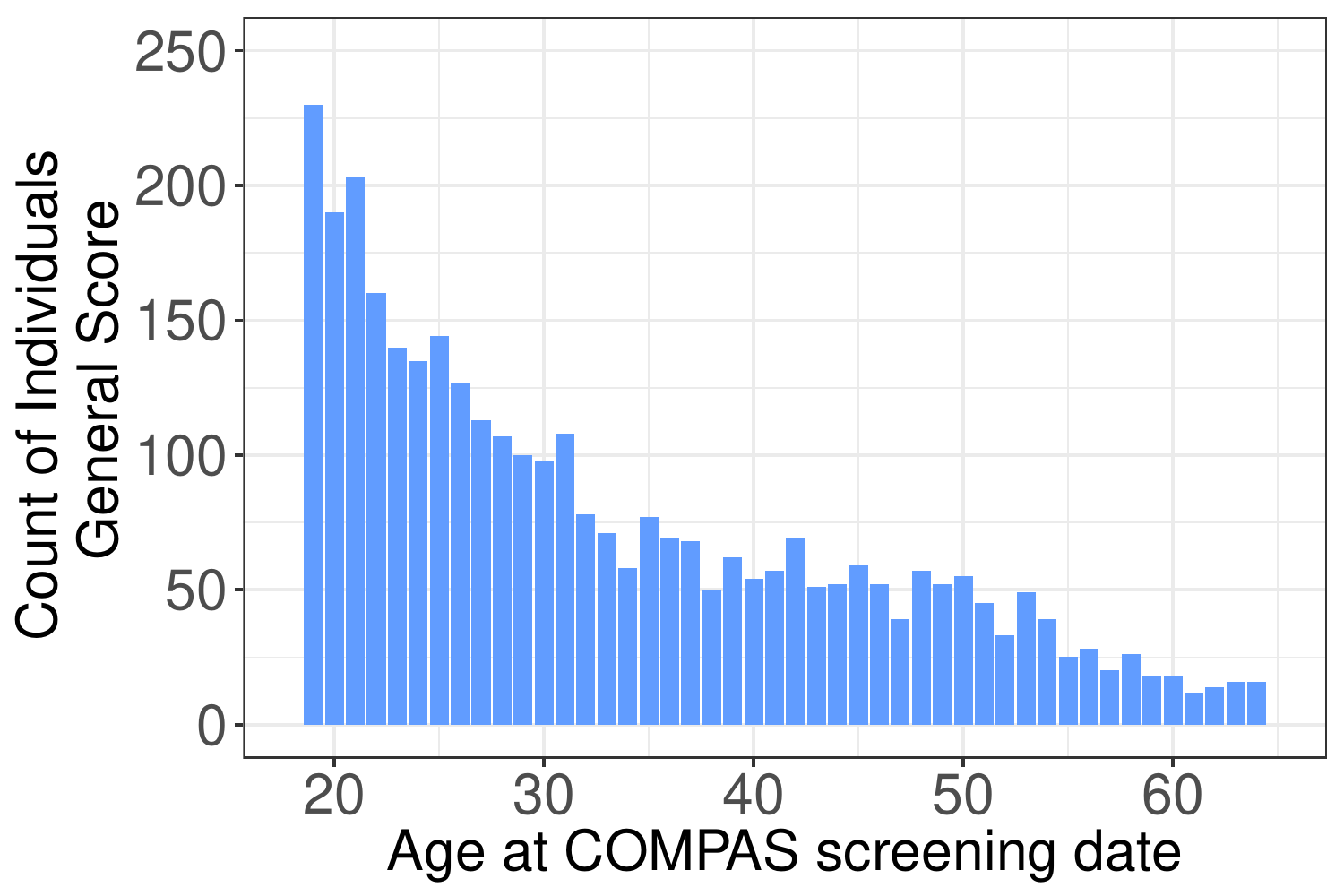}
  \caption{}
\end{subfigure}
\caption{ Age vs the count of individuals who satisfy the following conditions: (1) Age equals age-at-first-arrest, (2) All zeros for the subscale inputs in our data, and that correspond to the particular COMPAS score we consider (i.e. Criminal Involvement Subscale for the general score; History of Violence Subscale and History of Noncompliance Subscale for the violence score).
 \label{fig:check_dat_assump}}
\end{figure}

As explained in Section \ref{sec:howcompasdependsonage}, we make the assumption that there exist individuals in our data with the lowest possible COMPAS score for each age.
We do not have all the inputs to the COMPAS score, but if the assumption holds for the inputs we have access to, it lends evidence that the Data Assumption holds generally. In Figure \ref{fig:check_dat_assump}, we show the counts of individuals who have current age equal to age-at-first-arrest and have all zero values for the COMPAS subscale inputs within our data. 

%%%%%%%%%%%%%%%%%
\subsubsection*{age-at-first-arrest}
\begin{figure}[htbp]
\centering
\begin{subfigure}{.48\columnwidth}
  \centering
  \includegraphics[width=1\columnwidth]{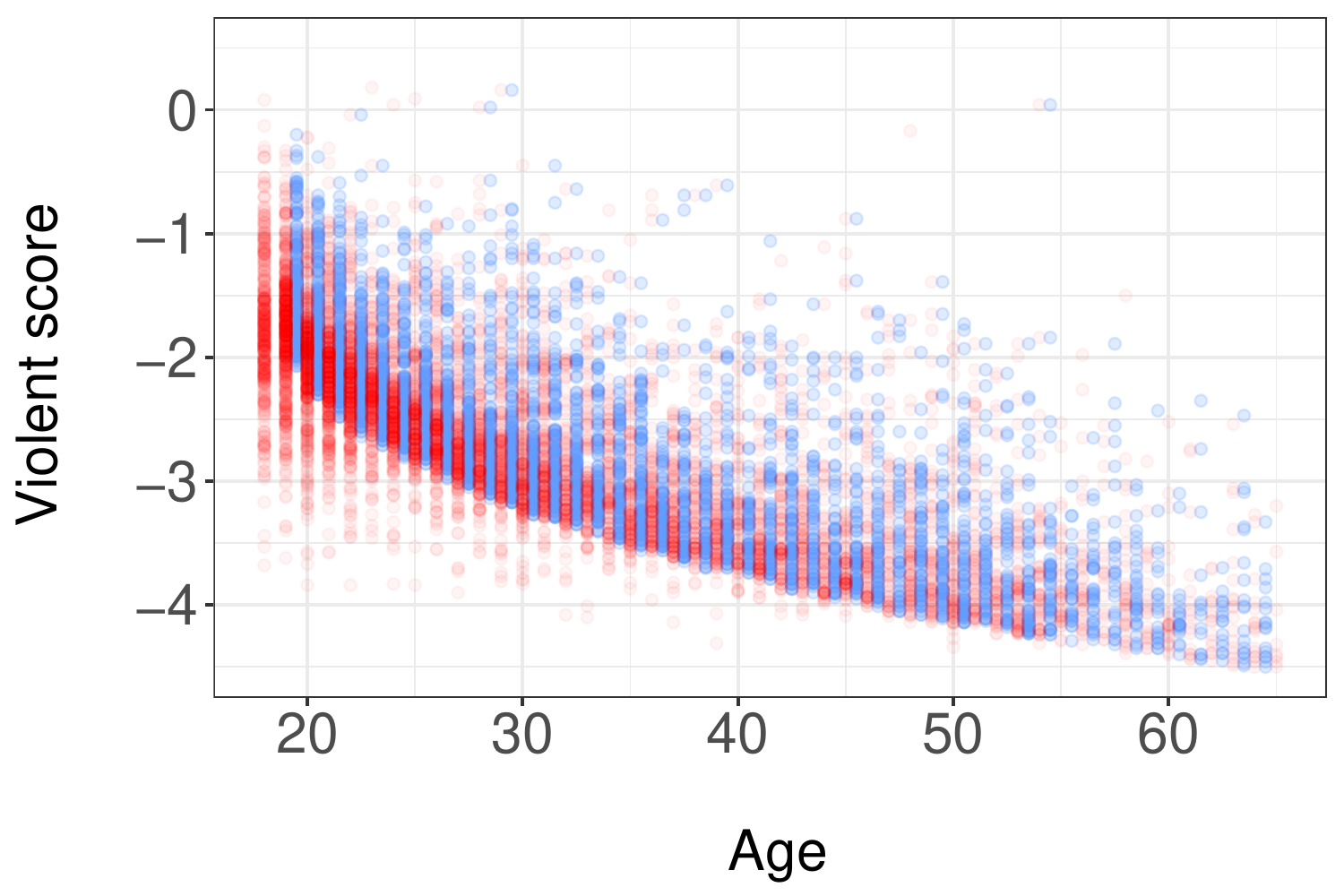}
  \caption{}
\end{subfigure} \hfill
\begin{subfigure}{.48\columnwidth}
  \centering
  \includegraphics[width=1\linewidth]{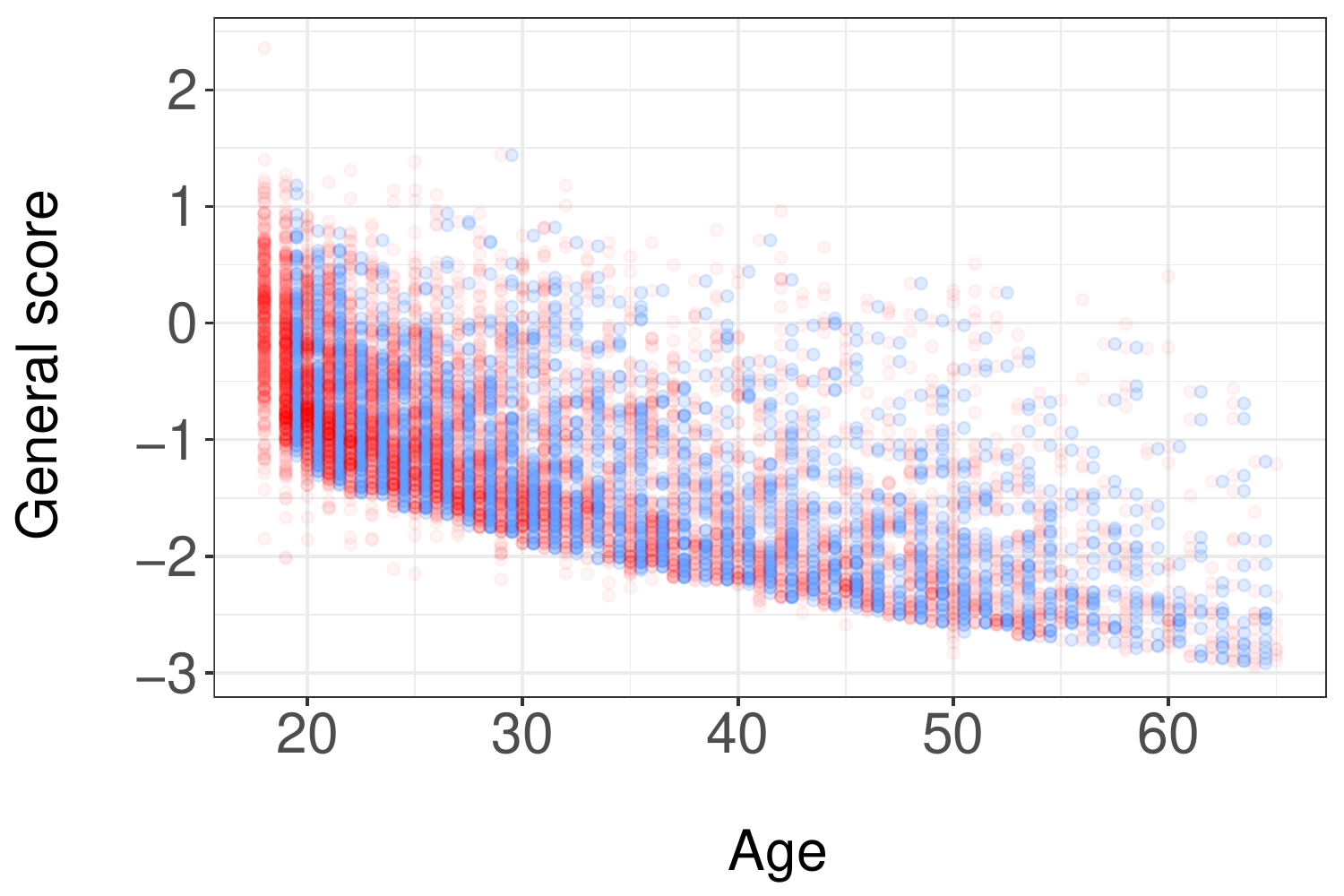}
  \caption{}
\end{subfigure}
\caption{ COMPAS raw score versus the age variables. age-at-first-arrest is depicted in red, while current age is depicted in blue and jittered to the right. There is no clear pattern for the lower bound with age-at-first-arrest, as there is for current age. Thus, age-at-first-arrest is not used to define an initial lower bound.
 \label{fig:age_first_offense}}
\end{figure}

The COMPAS lower bounds $f_{\textrm{age}}$ and $ f_{\textrm{viol age}}$  are defined by the current age at which the COMPAS score is computed, not the age-at-first-arrest. We chose to begin the analysis with current age because (1) the relationship between current age and the COMPAS score is the clearest in the data, and (2) according to the COMPAS documentation, the age variables are the only variables that have a linear relationship with the score. 
%This because individuals for whom age-at-first-arrest does not equal age (at current offense) are not the individuals needed for defining either $f_{\textrm{age}}$ or $ f_{\textrm{viol age}}$. 
Figure \ref{fig:age_first_offense} shows that there is not a smooth lower bound for COMPAS vs. age-at-first-arrest as there was with the current age. 

%%%%%%%%%%%%%%%%%%%%%%%%%%%%%%%%%
\subsubsection*{Individuals on the lower bound having the lowest possible COMPAS scores for their age}

The functions $f_{\textrm{age}}$ and $ f_{\textrm{viol age}}$ are defined by individuals who have zero values for the subscale components we can compute, and lie close to the age spline (i.e. were not deemed age outliers)---in other words, the individuals who \textit{could} satisfy the Data Assumption. While there is no way for us to know if these individuals truly satisfy the Data Assumption, in this section we explain why we believe these individuals actually do satisfy the assumption.

\begin{figure}[htbp]
\centering
\begin{subfigure}{.48\columnwidth}
  \centering
  \includegraphics[width=1\columnwidth]{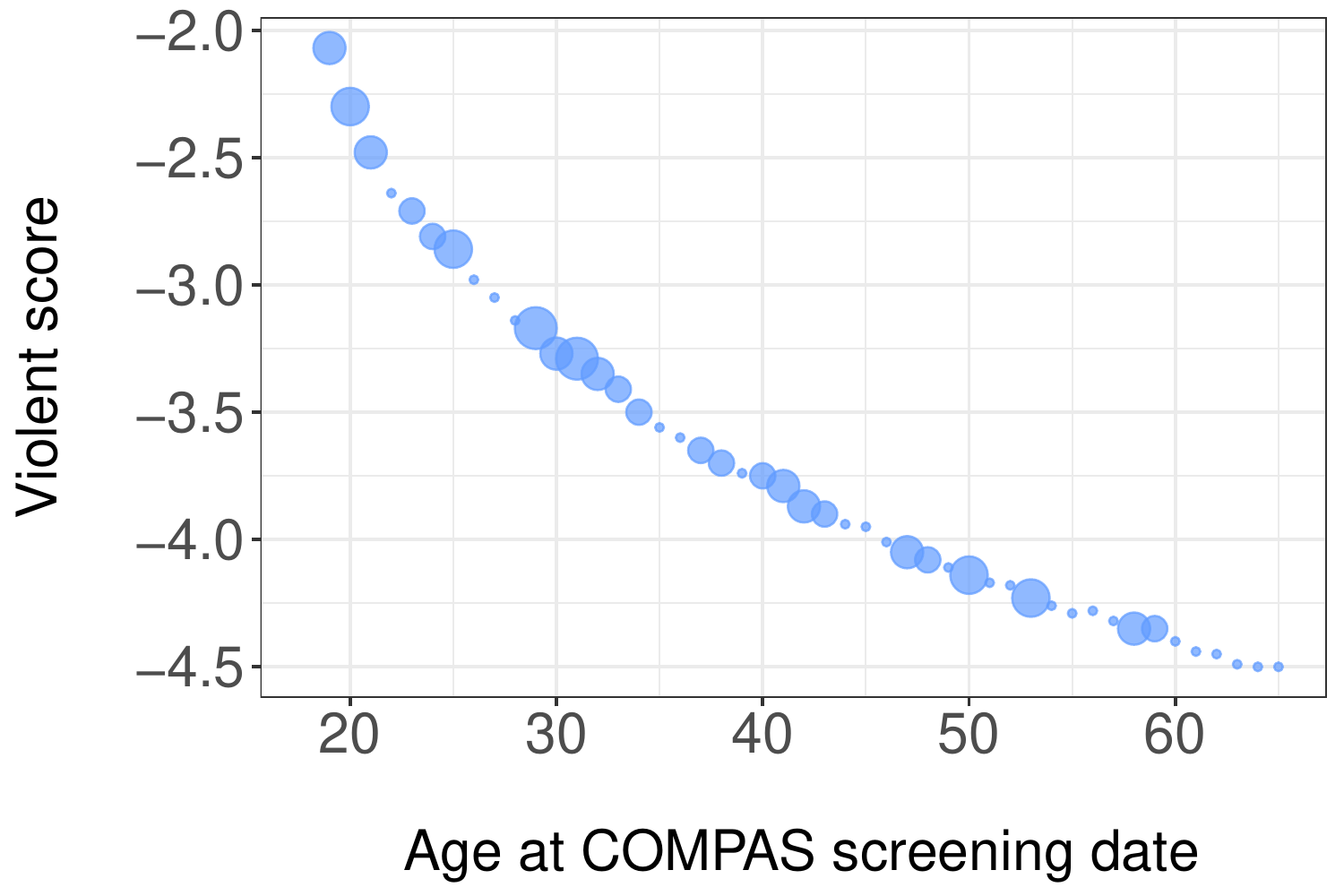}
  \caption{Lowest violent COMPAS raw score for each age, for individuals with no criminal history and no history of non-compliance. The smallest balls represent only one individual, the larger balls represent 7 individuals with identical COMPAS raw scores. 83 individuals are represented in this plot.}
  \label{fig:bubble_viol}
\end{subfigure} \hfill
\begin{subfigure}{.48\columnwidth}
  \centering
  \includegraphics[width=1\linewidth]{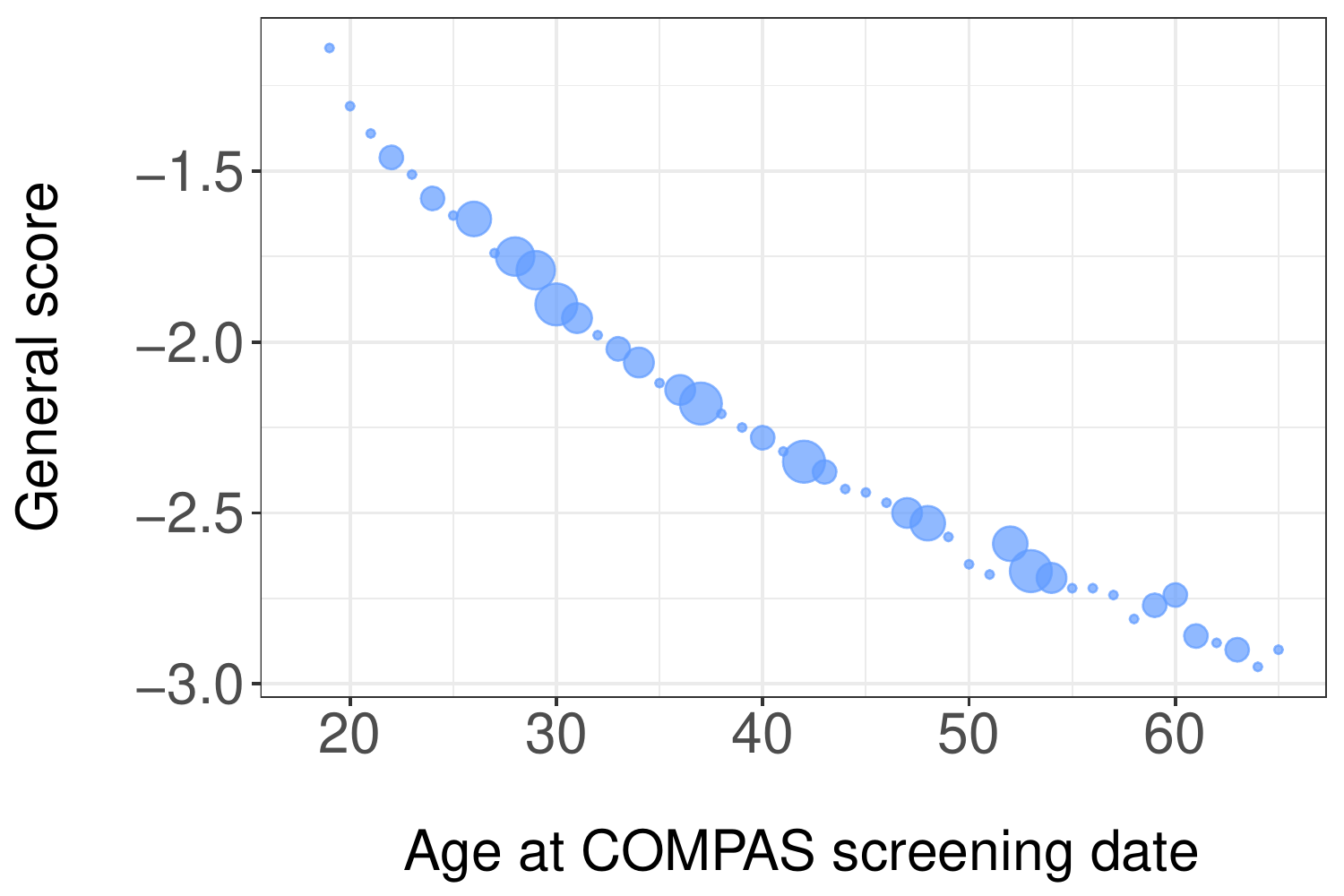}
  \caption{Lowest general COMPAS raw score for each age, for individuals with no criminal involvement. The smallest balls represent only one individual, the larger balls represent 6 individuals with identical COMPAS raw scores. 103 individuals are represented in this plot.}
  \label{fig:bubble_recid}
\end{subfigure}
\caption{ 
 \label{fig:bubble}}
\end{figure}

\begin{figure}[htbp]
\centering
\begin{subfigure}{.48\columnwidth}
  \centering
  \includegraphics[width=1\columnwidth]{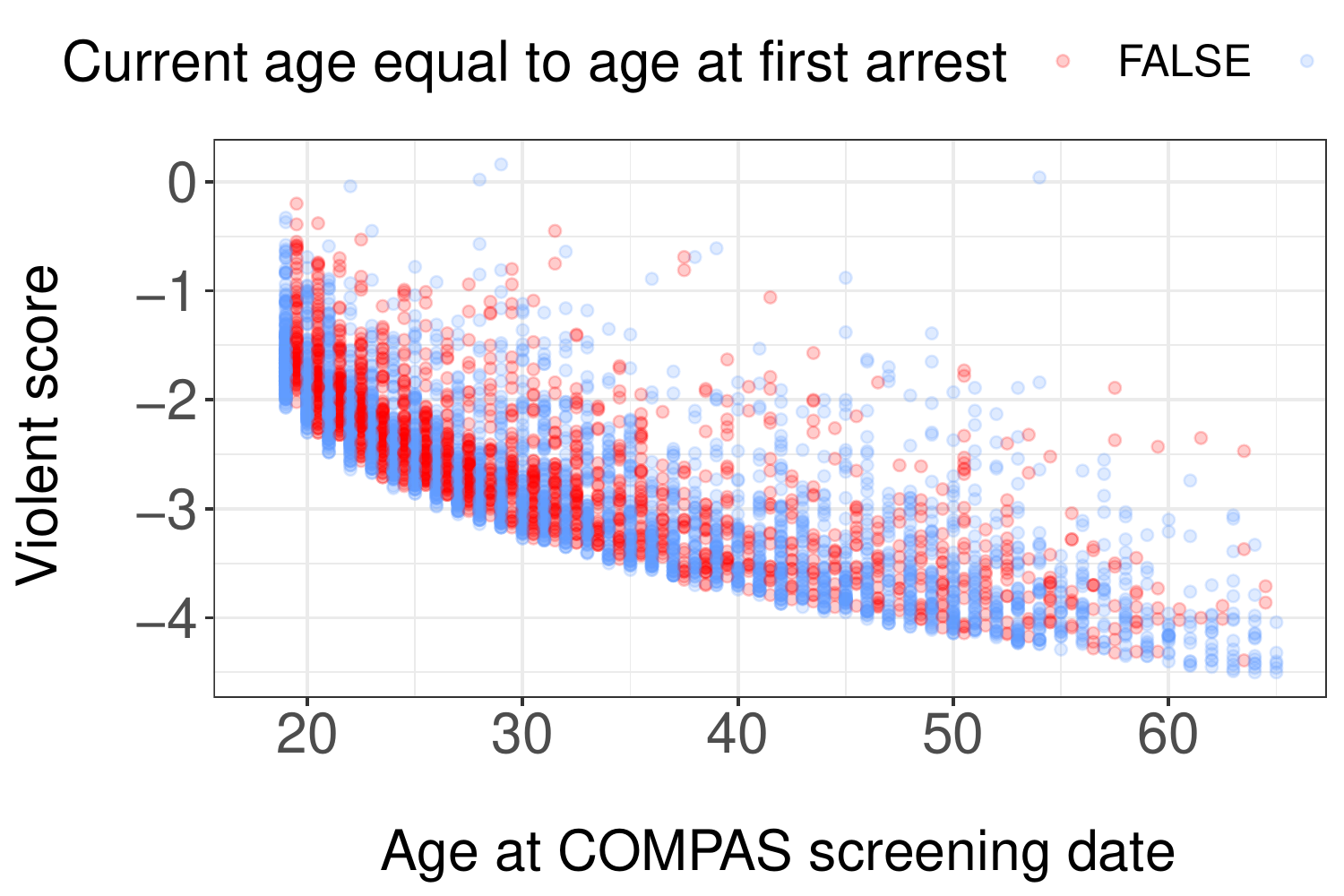}
  \caption{}
\end{subfigure} \hfill
\begin{subfigure}{.48\columnwidth}
  \centering
  \includegraphics[width=1\linewidth]{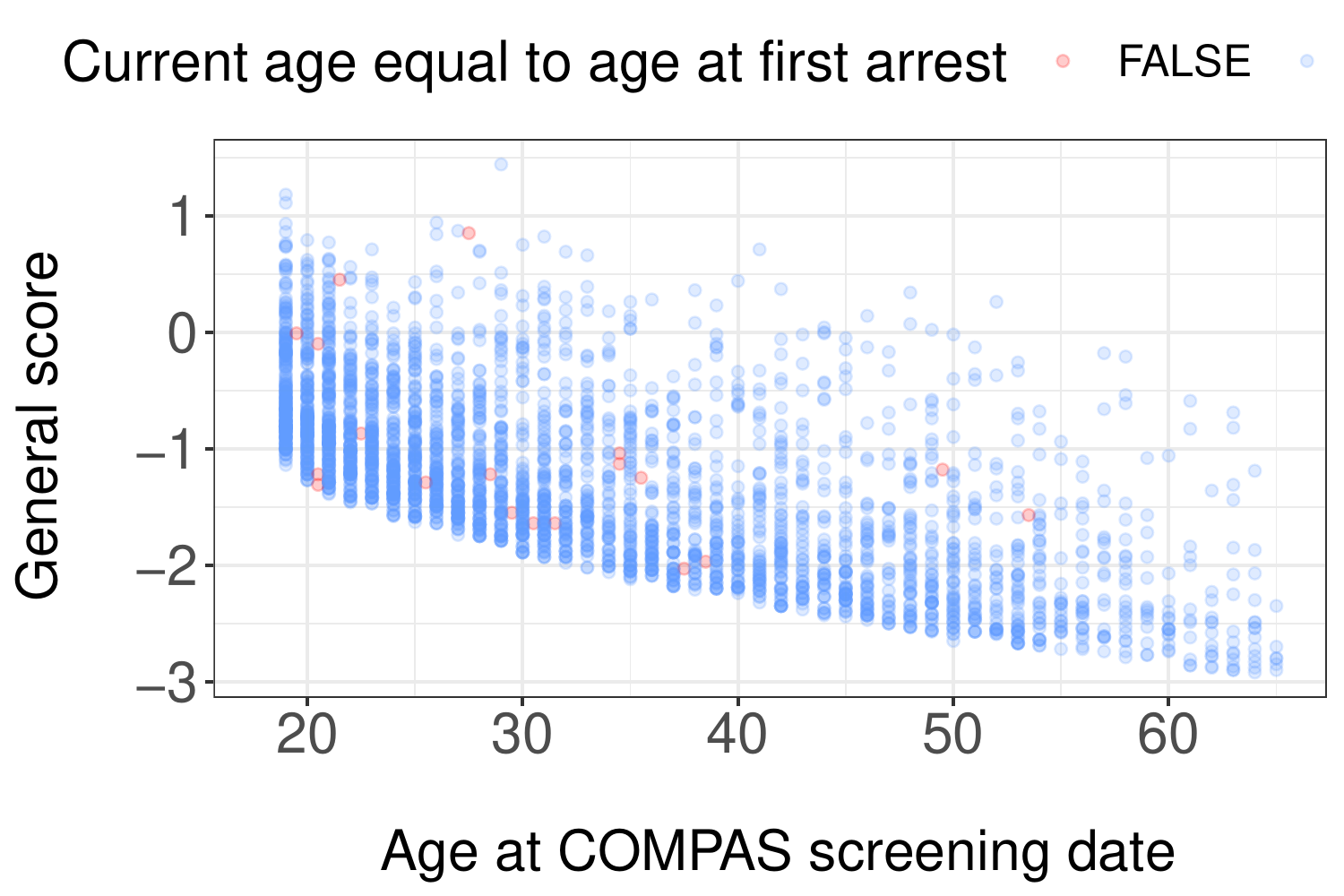}
  \caption{}
\end{subfigure}
\caption{ COMPAS raw score versus current age. Individuals such that current age is equal to age-at-first-arrest are shown in blue, whereas others for which current age is not equal to age-at-first-arrest are shown in red. The balls are semi-transparent, so purple balls consist of overlapping blue and red balls. We define the lower boundary for the Violence Score using individuals that have the additional constraint that they cannot have  violence history or non-compliance history; for the General Score, we use the constraint that individuals cannot have criminal involvement history. Note that individuals on the lower bound are mostly blue---that is, individuals have age equal to age at first arrest. Others tend to have higher COMPAS scores.
 \label{fig:fireplot}}
\end{figure}

Figure \ref{fig:bubble} shows raw COMPAS scores for these individuals who define the age splines. For many ages, there are several individuals whose COMPAS raw scores have identical values.  The number of such individuals for each age is shown by the size of the ball in the figure. Their current age is usually equal to their age-at-first-arrest. Figure \ref{fig:fireplot} shows this, where individuals with current age equal to age-at-first-arrest are plotted in blue, and others are in red.

%%%%%%%%%%%%%%%

It is possible that these individuals have nonzero values for the unobserved subscale components, which would imply that the true age functions could lie below $f_{\textrm{age}}$ and $ f_{\textrm{viol age}}$. However, we believe that Figure \ref{fig:bubble} -- which shows that that for many age values there are multiple individuals with exactly the same COMPAS raw score on the lower bound of our data -- combined with the smoothness of the lower bound, provides compelling evidence that all these individuals actually have zero values for the unobserved subscale components. Some alternative hypotheses and the reasons why we find them unlikely, are outlined below:
\begin{itemize}
   \item Some individuals on the lower bound have zeros for unobserved subscale components, others have nonzero values for the unobserved subscale components. We find this unlikely for two reasons. Since age and age-at-first-arrest are the only inputs with negative contribution to the COMPAS score, if we fix age, COMPAS scores for the individuals with positive values for unobserved subscale components should be higher than the lowest possible COMPAS score. Thus, for these individual to lie on the lower bound we observe, there cannot be \textit{any} individuals in the data who have the lowest possible COMPAS score for that age. Second, this would have to occur in a systematic way, to create the smooth, nonlinear lower bound we observe.
   
    \item All individuals on the lower bound have nonzero values for unobserved inputs because some of these inputs contribute negatively to the COMPAS score, thereby negating the positive contribution from other inputs, allowing the individual to have the lowest possible COMPAS score. We find this unlikely because according to the COMPAS documentation, age and age-at-first-arrest are the only variables that contribute negatively to the COMPAS score. As discussed above, age-at-first-arrest equals age for most of the individuals on the lower bound, so there is no contribution from age-at-first-arrest. Though we claim the COMPAS documentation is not completely accurate, we believe it is reliable in this respect because it would be unintuitive for the other inputs to contribute negatively to the COMPAS scores. 
    
    \item Some individuals on the lower bound have nonzero values for unobserved inputs, but manual overrides occurred for all of these individuals so that the only contribution to their score was age. This is possible, but implies that the only factor influencing their COMPAS scores was age, which then agrees with our hypothesis.

\end{itemize}

For these reasons, we believe that these individuals have the lowest possible COMPAS scores for their age and lie on the true age functions.

%%%%%%%%%%%%%%%%%%%%%%%%%%%
\subsection*{An alternative explanation of the nonlinearity in age}

In this section, we briefly consider an alternative explanation for the nonlinearity in age. As observed in Section \ref{sec:ageonly}, there are a greater proportion of individuals at lower ages than at higher ages in our data. It has been proposed that if age follows a Gaussian distribution with respect to the COMPAS score, it is possible that the nonlinear behavior we observe in the lower bound of individuals' COMPAS raw scores vs current age is induced by more extreme values appearing when we have more samples. However, exploratory analysis mitigates this concern. 

\begin{figure}[htbp]
\centering
\begin{subfigure}{.48\columnwidth}
  \centering
  \includegraphics[width=1\columnwidth]{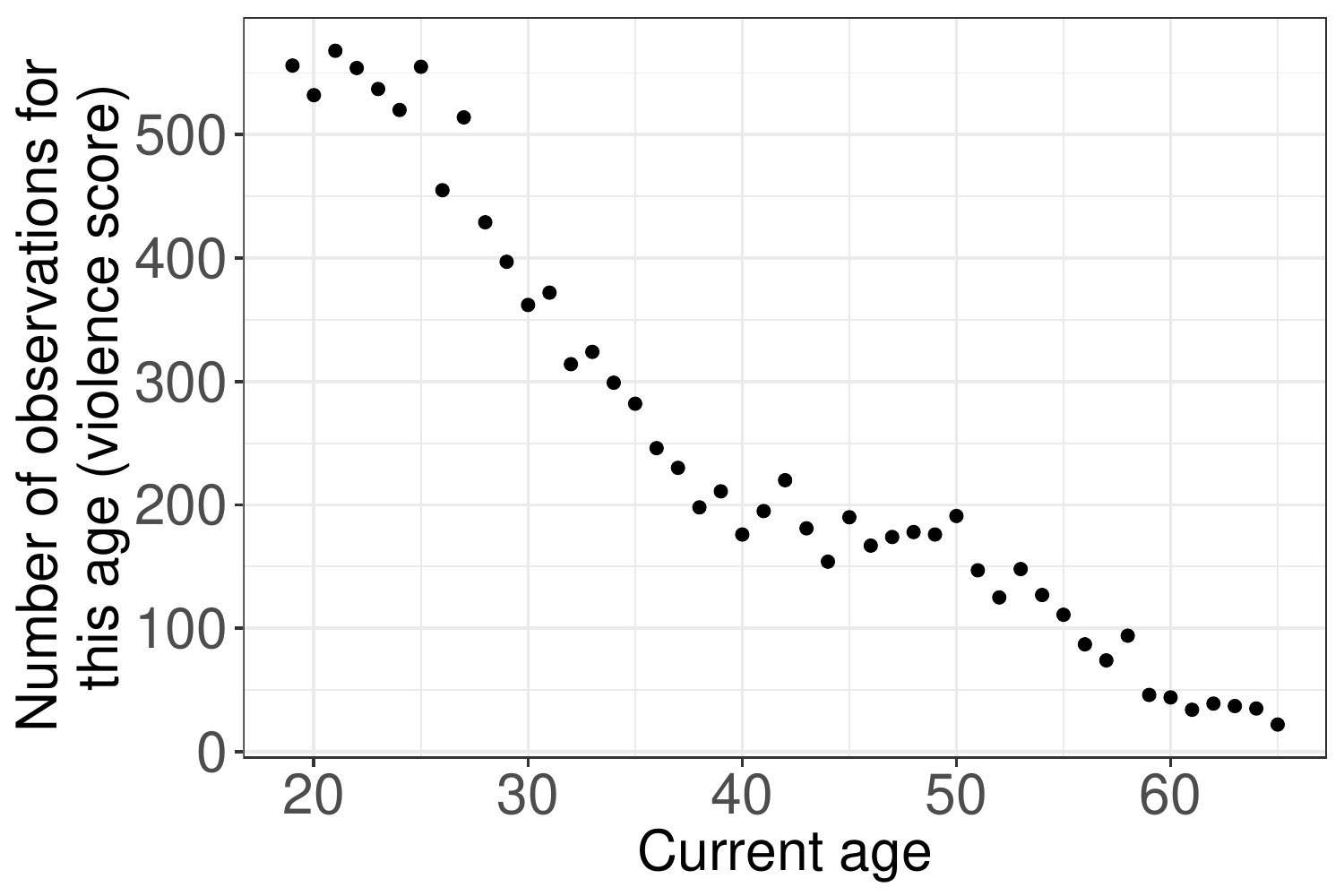}
  \caption{}
\end{subfigure} \hfill
\begin{subfigure}{.48\columnwidth}
  \centering
  \includegraphics[width=1\linewidth]{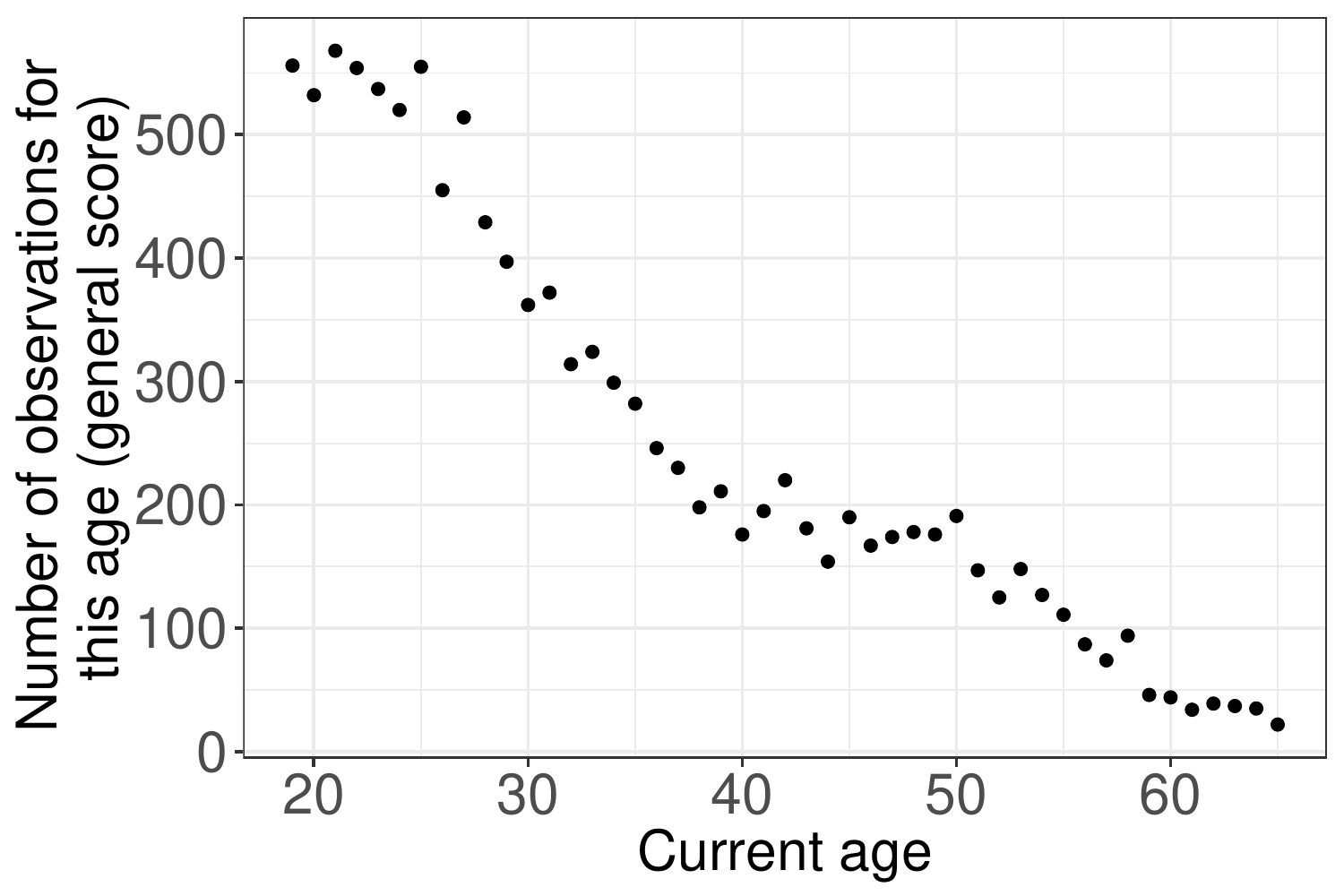}
  \caption{}
\end{subfigure}
\caption{ 
 \label{fig:num-inds-per-age}}
\end{figure}

\begin{figure}[htbp]
\centering
\begin{subfigure}{.48\columnwidth}
  \centering
  \includegraphics[width=1\columnwidth]{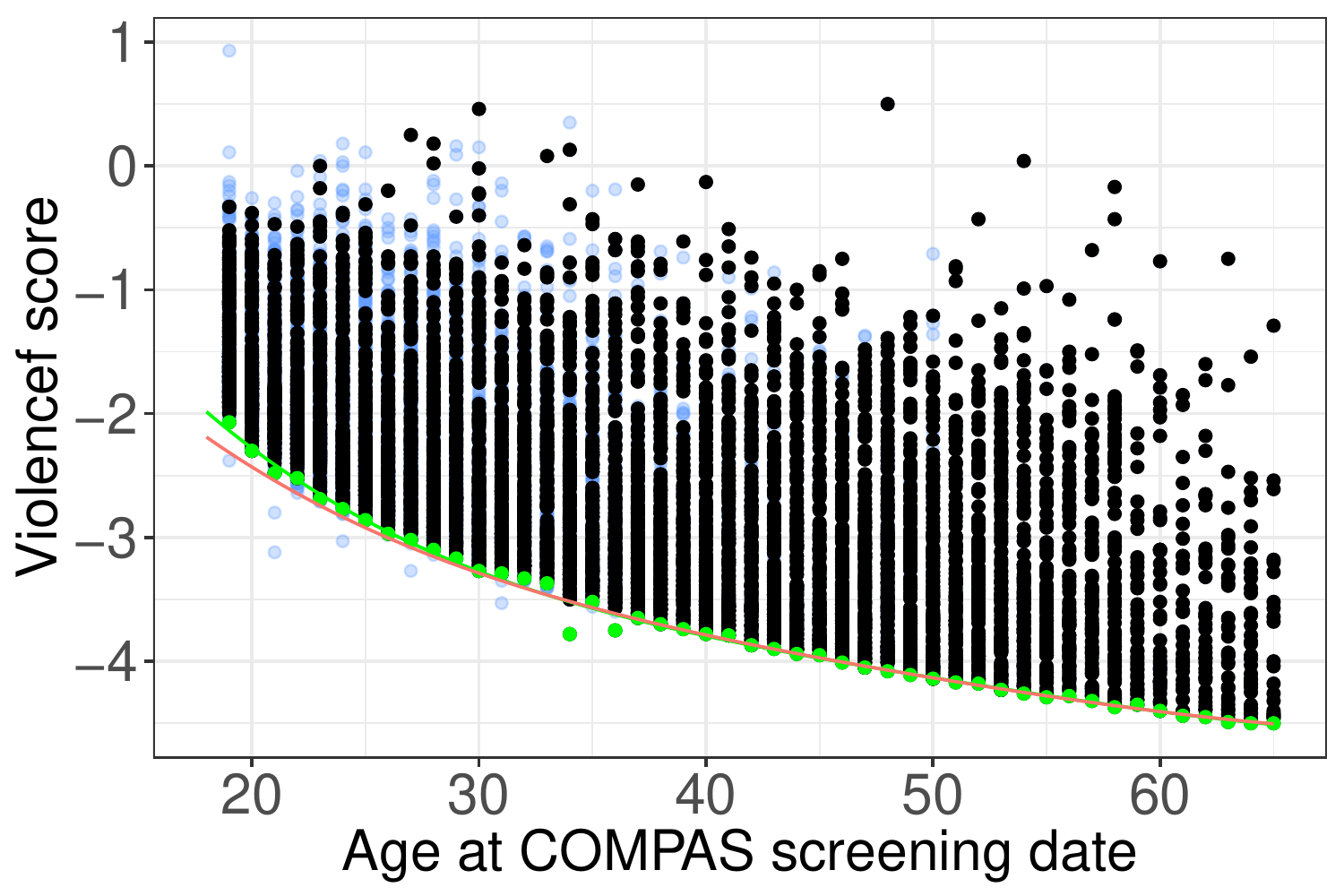}
  \caption{}
\end{subfigure} \hfill
\begin{subfigure}{.48\columnwidth}
  \centering
  \includegraphics[width=1\linewidth]{age_samp_viol.pdf}
  \caption{}
\end{subfigure}
\caption{ COMPAS raw score versus the age variables. The initial age polynomials fitted using the complete data are in red. The age polynomials fitted using a sample of $n = \min(150,\text{\textit{number individuals per age}})$ are in green. The two bounds are extremely close, mitigating the concern that the nonlinearity is due to the number of observations per age group.
 \label{fig:age-sampling}}
\end{figure}

Figure \ref{fig:num-inds-per-age} shows the number of individuals per age group. As Figure \ref{fig:age-sampling} shows, sampling $\min(150,\text{\text{number of individuals per age}}) $ for each age group does not alter the nonlinearity of the lower bound. Moreover, age splines fit using only the sampled data are very close to the original age splines.

%%%%%%%%%%%%%%%%%%%%%%%%%%%%%%%%%
%\subsubsection*{Quantile regression}
%In this subsection we warn the reader against a type of analysis that should not be done, and discuss why. 
%One might try to show that the age polynomial is independent of the remainder term COMPAS $- f_{\textrm{age}}$ (or COMPAS $- f_{\textrm{viol age}}$ for the violence remainder term). However, age is \textit{not} independent of the remainder because of the correlation of age with criminal history. In an attempt to show this independence (which will fail), one might try to plot the quantiles of COMPAS for each age group. If these quantiles are evenly spaced, it would show independence. As Figure~\ref{fig:quantiles} shows, the lines are not evenly spaced. Thus, this type of analysis is not helpful.
%\begin{figure}[htbp]
%\centering
%\begin{subfigure}{.48\textwidth}
%  \centering
%  \includegraphics[width=1\linewidth]{quantiles_general.pdf}
%  \caption{}
  %\label{fig:sub1}
%\end{subfigure} \hfill
%\begin{subfigure}{.48\textwidth}
%  \centering
%  \includegraphics[width=1\linewidth]{quantiles_violent.pdf}
%  \caption{}
  %\label{fig:sub2}
%\end{subfigure}
%\caption{An experiment that cannot work because age is not independent from criminal history.\label{fig:quantiles}}
%\end{figure}
%%%%%%%%%%%%%%%%%%%%%%%%
%%%%%%%%%%%%%%%%%%%%%%%%%%%%%%%%%%%
\subsection*{Logistic Regression}

We attempt to replicate ProPublica's logistic regression of the COMPAS score category (Medium or High Risk vs. Low Risk) on various features, including race. 
Coefficient estimates and standard errors are shown in Table \ref{table:propubregression}. Since recidivism (the outcome for our recidivism prediction models) is used as a covariate in ProPublica's analysis, we exclude any observation for which there is less than two years of data beyond the screening date. Note that if 2-year recidivism is used in ProPublica's model, it is using information that by definition is not available at the time that the COMPAS score is calculated.

\begin{table}
\centering
\begin{tabular}{rccccc}
\cline{2-6}
\multicolumn{1}{l}{} & \multicolumn{2}{c}{Our Results} & \multicolumn{1}{l}{} & \multicolumn{2}{c}{ProPublica's Results} \\ \cline{2-6} 
\multicolumn{1}{l}{} & Estimate     & Standard Error   &                      & Estimate         & Standard Error        \\ \cline{2-6} 
Female               & 0.123     & 0.085            &                      & 0.221***         & 0.080                 \\
Age: Greater than 45 & -1.489***    & 0.129            &                      & -1.356***        & 0.099                 \\
Age: Less than 25    & 1.445***     & 0.071            &                      & 1.308***         & 0.076                 \\
Black                & 0.521***     & 0.072            &                      & 0.477***         & 0.069                 \\
Asian                & -0.271       & 0.503            &                      & -0.254***        & 0.478                 \\
Hispanic             & -0.301*      & 0.130            &                      & -0.428***        & 0.128                 \\
Native American      & 0.390        & 0.678            &                      & 1.394*           & 0.766                 \\
Other                & -0.713***    & 0.159            &                      & -0.826***        & 0.162                 \\
Number of Priors     & 0.155***     & 0.006            &                      & 0.269***         & 0.011                 \\
Misdemeanor          & -0.464***    & 0.069            &                      & -0.311***        & 0.067                 \\
Two year Recidivism  & 0.491***     & 0.068            &                      & 0.686***         & 0.064                 \\
Constant             & -1.593***    & 0.082            &                      & -1.526***        & 0.079                
\end{tabular}
\caption{Logistic regression coefficient estimates and standard errors computed in a similar way to ProPublica. The significance levels are not valid, since the model assumptions of linearity are broken. *$p<0.1$; **$p<0.05$; ***$p<0.01$. Our results are based on 5759 observations while ProPublica's results are based on 6,172 observations. \label{table:propubregression}}
\end{table}

\subsection*{Data processing}
Our data includes the same raw data collected by ProPublica, which includes COMPAS scores for all individuals who were scored in 2013 and 2014, obtained from the Broward County Sheriff’s Office. There are 18,610 individuals, but we follow ProPublica in examining only the 11,757 of these records which were assessed at the pretrial stage. We also used public criminal records from the Broward County Clerk's Office to obtain the events/documents and disposition for each case, which we used in our analysis to infer probation events.

In their analysis \cite{LarsonMaKiAn16,AngwinLaMaKi16}, ProPublica processed the raw data, which includes charge, arrest, prison, and jail information, into features aggregated by person, like the number of priors or whether or not a new charge occurred within two years. We too process the raw data into features, partly to ensure the quality of the features and partly to create new features as defined by the components of the COMPAS subscales (see Tables \ref{table:features_viol}-\ref{table:features_drug}). Note that while ProPublica publishes the code for their analysis and the raw data, they do not publish the code for processing the raw data. Thus we did that from scratch. 

The \textit{screening date} is the date on which the COMPAS score was calculated.

\begin{itemize}
\item Our features correspond to an individual on a particular screening date. If a person has multiple screening dates, we compute the features for each screening date, such that the set of events for calculating features for earlier screening dates is included in the set of events for later screening dates. 

\item On occasion, an individual will have multiple COMPAS scores calculated on the same date. There appears to be no information distinguishing these scores other than their identification number. We take the scores with the larger identification number. 

\item Any charge with degree ``(0)'' seems to be a very minor offense, so we exclude these charges. All other charge degrees are included, meaning charge degrees other than felonies and misdemeanors are included. 

\item Some components of the Violence Subscale require classifying the type of each offense (\textit{e.g.}, whether or not it is a weapons offense). We infer this from the statute number, most of which correspond to statute numbers from the Florida state crime code. 

\item The raw data includes arrest data as well as charge data. Because the arrest data does not include the statute, which is necessary for the Violence Subscale, we use the charge data and not the arrest data throughout the analysis. While the COMPAS subscales appear to be based on arrest data, we believe the charge data should provide similar results. 

\item For each person on each COMPAS screening date, we identify the offense --- which we call the \textit{current offense} --- that we believe triggered the COMPAS screening. The \textit{current offense date} is the date of the most recent charge that occurred on or before the COMPAS screening date. Any charge that occurred on the current offense date is part of the current offense. In some cases, there is no prior charge that occurred near the COMPAS screening date, suggesting charges may be missing from the dataset. For this reason we consider charges that occurred within 30 days of the screening date for computing the current offense. If there are no charges in this range, we say the current offense is missing. For any part of our analysis that requires criminal history, we exclude observations with missing current offenses. All components of the COMPAS subscales that we compute are based on data that occurred prior to (not including) the current offense date, which is consistent with how the COMPAS score is calculated according to \cite{compas}. 

\item The events/documents data includes a number of events (\textit{e.g.}, ``File Affidavit Of Defense'' or ``File Order Dismissing Appeal'') related to each case, and thus to each person. To determine how many prior offenses occurred while on probation, or if the current offense occurred while on probation, we define a list of event descriptions indicating that an individual was taken on or off probation. Unfortunately, there appear to be missing events, as individuals often have consecutive ``On'' or consecutive ``Off'' events (e.g., two ``On" events in a row, without an ``Off" in between). In these cases, or if the first event is an ``Off'' event or the last event is an ``On'' event, we define two thresholds, $t_{\textrm{on}}$ and $t_{\textrm{off}}$. If an offense occurred within $t_{\textrm{on}}$ days after an ``On'' event or $t_{\textrm{off}}$ days before an ``Off'' event, we count the offense as occurring while on probation. We set $t_{\textrm{on}}$ to 365 and $t_{\textrm{off}}$ to 30.
On the other hand, the ``number of times on probation" feature is just the count of ``On'' events and the ``number of times the probation was revoked" feature is just the count of ``File order of Revocation of Probation'' event descriptions (\textit{i.e.}, there is no logic for inferring missing probation events for these two features).

\item Age is defined as the age in years, rounded down to the nearest integer, on the COMPAS screening date.

\item Recidivism is defined as any charge that occurred within two years of the COMPAS screening date. For any part of our analysis that requires recidivism, we use only observations for which we have two years of subsequent data.

\item A juvenile charge is defined as an offense that occurred prior to the defendant's 18th birthday. 

\end{itemize}

%%%%%%%%%%%%%%%%%%%%%%%%%
\subsection*{Machine learning implementation}

Here we discuss the implementation of the various machine learning methods used in this paper.
To predict the COMPAS general and violence raw score remainders (Tables \ref{table:withandwithoutage}, \ref{table:withandwithout}, and \ref{table:withandwithout_violent}), we use a linear regression (base \texttt{R}), random forests (\texttt{randomForest} package), Extreme Gradient Boosting (\texttt{xgboost} package), and SVM (\texttt{e1071} package). To clarify, we predict the COMPAS raw scores (not the decile scores, since these are computed by comparing the raw scores to a normalization group) after subtracting the age polynomials ($f_{\textrm{age}}$ for the general raw score and $f_{\textrm{viol age}}$ for the violence raw score). For XGBoost and SVM we select hyperparameters by performing 5-fold cross validation on a grid of hyperparameters and then re-train the method on the set of hyperparameters with the smallest cross validation error. For random forest we use the default selection of hyperparameters. For the COMPAS general raw score remainder, we use the available Criminal Involvement Subscale features (Table \ref{table:features_crim}), while for the COMPAS violence raw score remainder, we use the available History of Violence Subscale and History of Noncompliance Subscale features listed in tables Tables \ref{table:features_viol} and \ref{table:features_noncomp}, respectively. For both types of COMPAS raw scores, we also use the age-at-first-arrest. Race and age at screening date may or may not be included as features, as indicated when the results are discussed. 
To predict general and violent two-year recidivism (Tables \ref{table:recid_withandwithout} and \ref{table:recid_withandwithout_violent}), we use the same methods, features, and cross validation technique as used to predict the raw COMPAS score remainders, except we adapt each method for classification instead of regression (for linear regression, we substitute logistic regression) and we include the current offense in the features. All code is written in \texttt{R} and is available on GitHub\footnote{\url{https://github.com/beauCoker/age_of_unfairness}}.

%%%%%%%%%%%%%%%%%%%%%%%%%

\subsection*{Subscale tables}

The features that compose the subscales used by COMPAS and that we use for prediction are listed in Tables \ref{table:features_viol}-\ref{table:features_drug}. The Criminal History, Substance Abuse, and Vocation/Education Subscales (Tables \ref{table:features_crim}, \ref{table:features_drug}, and \ref{table:features_voca}, respectively) are inputs to the COMPAS general recidivism score, while the History of Violence, History of Noncompliance, and Vocation/Education Subscales (Tables \ref{table:features_viol}, \ref{table:features_noncomp}, and \ref{table:features_voca}, respectively) are inputs to the COMPAS violent recidivism score.

%%%%%%%%%%%%%%%

%%%History of Violence Subscale Items %%%
\begin{table}
\centering
\caption{History of Violence Subscale. We compute the components in bold font. We do not have the data to compute the other components. The feature for family violent arrests is always 0 so it is not useful for prediction. We classify a charge as family violence if the statute is 741.28, which corresponds to the definition of domestic violence in the Florida crime code. In our dataset there were no instances of this statute. 
\label{table:features_viol}}
\begin{tabular}{@{}ll@{}}
\toprule
Subscale Items                                                                                                                            & Values       \\ \midrule
\textbf{Prior juvenile felony offense arrests}                                                                                            & \textbf{0,1,2+}       \\
\textbf{\begin{tabular}[c]{@{}l@{}}Prior violent felony property offense arrests\end{tabular}}                                        & \textbf{0,1,2,3,4,5+} \\
\textbf{\begin{tabular}[c]{@{}l@{}}Prior murder/voluntary manslaughter   arrests\end{tabular}}                                          & \textbf{0,1,2,3+}     \\
\textbf{\begin{tabular}[c]{@{}l@{}}Prior felony assault offense arrests\\   (excluding murder, sex, or domestic violence)\end{tabular}}   & \textbf{0,1,2,3+}     \\
\textbf{\begin{tabular}[c]{@{}l@{}}Prior misdemeanor assault offense arrests\\   (excluding murder, sex, domestic violence)\end{tabular}} & \textbf{0,1,2,3+}     \\
\textbf{Prior family violence arrests}                                                                                                    & \textbf{0,1,2,3+}     \\
\textbf{Prior sex offense arrests}                                                                                                        & \textbf{0,1,2,3+}     \\
\textbf{Prior weapons offense arrest}                                                                                                     & \textbf{0,1,2,3+}     \\
\begin{tabular}[c]{@{}l@{}}Disciplinary infractions for\\   fighting/threatening other inmates/staff\end{tabular}                         & Yes/No       \\ \bottomrule
\end{tabular}
\end{table}
\quad

%%%History of Noncompliance Subscale%%%
\begin{table}
\centering
\caption{History of Noncompliance Subscale. We compute the components in bold font. We do not have the data to compute the other components.
\label{table:features_noncomp}}
\begin{tabular}{@{}ll@{}}
\toprule
Subscale Items                                                                                                 & Values                        \\ \midrule
\textbf{\begin{tabular}[c]{@{}l@{}}On probation or parole at time of current   offense*\end{tabular}}        & \textbf{Probation/Parole/Both/Neither*} \\
Number of parole violations                                                                                    & 0,1,2,3,4,5+                  \\
\begin{tabular}[c]{@{}l@{}}Number of times person has been returned\\   to prison while on parole\end{tabular} & 0,1,2,3,4,5+                  \\
\textbf{\begin{tabular}[c]{@{}l@{}}Number of new charge/arrests while on\\   probation\end{tabular}}           & \textbf{0,1,2,3,4,5+}                  \\
\textbf{\begin{tabular}[c]{@{}l@{}}Number of probation   violations/revocations\end{tabular}}                & \textbf{0,1,2,3,4,5+}    \\ 
\bottomrule
\multicolumn{2}{r}{{* = Only ``On Probation'' and ``Not On Probation'' computed. }}

\end{tabular}
\end{table}

%%%Criminal Involvement Subscale%%%
\begin{table}[]
\centering
\caption{Criminal Involvement Subscale. We computed all the components. To compute the number of arrests component, we interpreted a charge as an arrest.
\label{table:features_crim}}
\begin{tabular}{ll}
\toprule
Subscale Items  & Values  \\
\midrule
\textbf{\begin{tabular}[c]{@{}l@{}}Number times offender has been arrested as adult/juvenile \\for criminal offense\end{tabular}} & \textbf{Any value accepted} \\
\textbf{\begin{tabular}[c]{@{}l@{}}Number times offender sentenced to jail for $\geq$30 days\end{tabular}}                         & \textbf{0,1,2,3,4,5+}       \\
\textbf{\begin{tabular}[c]{@{}l@{}}Number of new commitments to state/federal prison\\ (include current)\end{tabular}}          & \textbf{0,1,2,3,4,5+}       \\
\textbf{\begin{tabular}[c]{@{}l@{}}Number of times person sentenced to probation as adult\end{tabular}}                         & \textbf{0,1,2,3,4,5+}  \\

\bottomrule
\end{tabular}
\end{table}

%%%Vocation/Education Subscale%%%
\begin{table}
\centering
\caption{Vocation/Education Subscale. We do not have the data to compute any of these components.
\label{table:features_voca}}
\begin{tabular}{@{}ll@{}}
\toprule
Subscale Items                                                                                                                           & Values                                                                   \\ \midrule
Completed high school diploma/GED                                                                                                        & Y/N                                                                      \\
Final grade completed in school                                                                                                          & \_\_\_                                                                   \\
Usual grades in high school                                                                                                              & A,B,C,D,E/F, did not attend                                              \\
Suspended/expelled from school                                                                                                           & Y/N                                                                      \\
Failed/repeated a grade level                                                                                                            & Y/N                                                                      \\
Currently have a job                                                                                                                     & Y/N                                                                      \\
\begin{tabular}[c]{@{}l@{}}Have a skill/trade/profession in which\\   you can find work\end{tabular}                                     & Y/N                                                                      \\
Can verify employer/school (if attending)                                                                                                & Y/N                                                                      \\
\begin{tabular}[c]{@{}l@{}}Amount of time worked or in school over\\   past 12 months\end{tabular}                                       & \begin{tabular}[c]{@{}l@{}}12 months full time,\\12 months part time,\\6$+$ months FT,\\0-6 months PT/FT\end{tabular} \\
\begin{tabular}[c]{@{}l@{}}Feel that you need more training in new\\   job or career skill\end{tabular}                                  & Y/N                                                                      \\
\begin{tabular}[c]{@{}l@{}}If you were to get (or have) a good job,\\   how would you rate your chance of being successful?\end{tabular} & Good, Fair, Poor                                                         \\
\begin{tabular}[c]{@{}l@{}}How hard is it for you to find a job\\   above minimum wage compared to others?\end{tabular}                  & \begin{tabular}[c]{@{}l@{}}Easier, Same,\\Harder, Much Harder\end{tabular}
\\ \bottomrule
\end{tabular}
\end{table}

%%%Substance Abuse Subscale%%%
\begin{table}[]
\centering
\caption{Substance Abuse Subscale. We do not have the data to compute any of these components.
\label{table:features_drug}}
\begin{tabular}{ll}
\toprule
Subscale Items  & Values  \\
\midrule
\begin{tabular}[c]{@{}l@{}}Do you think your current/past legal\\   problems are partly because of alcohol or drugs?\end{tabular} & Y/N    \\
\begin{tabular}[c]{@{}l@{}}Were you using alcohol when arrested for\\   your current offense?\end{tabular}                        & Y/N    \\
\begin{tabular}[c]{@{}l@{}}Were you using drugs when arrested for\\   your current offense?\end{tabular}                          & Y/N    \\
\begin{tabular}[c]{@{}l@{}}Are you currently in formal treatment for\\   alcohol/drugs?\end{tabular}                              & Y/N    \\
\begin{tabular}[c]{@{}l@{}}Have you ever been in formal treatment\\   for alcohol/drugs?\end{tabular}                             & Y/N    \\
\begin{tabular}[c]{@{}l@{}}Do you think you would benefit from\\   getting treatment for alcohol?\end{tabular}                    & Y/N    \\
\begin{tabular}[c]{@{}l@{}}Do you think you would benefit from\\   getting treatment for drugs?\end{tabular}                      & Y/N    \\
\begin{tabular}[c]{@{}l@{}}Did you use heroin, cocaine, crack, or\\   methamphetamines as a juvenile?\end{tabular}                & Y/N   \\
\bottomrule
\end{tabular}
\end{table}

%% file: tableindividuals.tex
%\onecolumn
\begin{footnotesize}
\begin{center}
\begin{longtable} {lllll}
\caption{Individuals whose COMPAS violence decile score is low (low-risk), but who have significant criminal histories. \# Priors counts the prior charges up to but not including the current offense (\textit{i.e.}, the offense we believe triggered the COMPAS score calculation), since this is how COMPAS counts prior offenses. However, the current offense may be included in Selected Prior Charges. Note that any Subsequent Crimes beyond the 2 year mark of the COMPAS score calculation or outside of Broward County may not be contained in our database. The notation ``(F/M,N)'' next to each charge gives the charge degree (F=Felony or M=Misdemeanor) and the number of instances of this charge (N).
\label{table:people}}   

\\ \hline Name                                                         & \begin{tabular}[c]{@{}l@{}}COMPAS\\ Violence Decile\end{tabular} & \begin{tabular}[c]{@{}l@{}}\#\\ Priors\end{tabular} & \begin{tabular}[c]{@{}l@{}}Selected\\ Prior Charges\end{tabular}                                                                                                                                  & \begin{tabular}[c]{@{}l@{}}Selected \\ Subsequent Charges\end{tabular}                                                                      \\ \hline \hline
\endfirsthead

\multicolumn{4}{c}%
{{\tablename\ \thetable{} --- Continued from previous page}} \\
\hline Name                                                         & \begin{tabular}[c]{@{}l@{}}COMPAS\\ Violence\\ Decile\end{tabular} & \begin{tabular}[c]{@{}l@{}}\#\\ Priors\end{tabular} & \begin{tabular}[c]{@{}l@{}}Selected\\ Prior Charges\end{tabular}                                                                                                                                  & \begin{tabular}[c]{@{}l@{}}Selected \\ Subsequent Charges\end{tabular}                                                                      \\ \hline \hline
\endhead

\multicolumn{5}{r}{{Continued on next page}} \\
%\hline \multicolumn{5}{l}{{$s$ = COMPAS Violence Decile, $n$ = Number of Prior Charges}}
\endfoot

%\hline \multicolumn{5}{l}{{$s$ = COMPAS Violence Decile, $n$ = Number of Prior Charges}}
\endlastfoot

%\begin{tabular}[c]{@{}l@{}}Martin \\ Owens\end{tabular}      & 1                                                                  & 0                                                   & Kidnapping (F,1)                                                                                                                                                                                  &                                                                                                                                             \\ \hline
\begin{tabular}[c]{@{}l@{}}Vilma \\ Dieppa\end{tabular}      & 1                                                                  & 4                                                   & \begin{tabular}[c]{@{}l@{}}Aggravated Battery (F,1), \\ Child Abuse (F,1), \\ Resist Officer w/Violence (F,1)\end{tabular}                                                                        &                                                                                                                                             \\ \hline
\begin{tabular}[c]{@{}l@{}}David \\ Selzer\end{tabular}      & 1                                                                  & 14                                                  & \begin{tabular}[c]{@{}l@{}}Battery on Law Enforc Officer (F,3), \\ Aggravated Assault W/Dead Weap (F,1), \\ Aggravated Battery (F,1), \\ Resist/obstruct Officer W/viol (F,1)\end{tabular}        &                                                                                                                                             \\ \hline
\begin{tabular}[c]{@{}l@{}}Berry \\ Sanders\end{tabular}     & 1                                                                  & 15                                                  & \begin{tabular}[c]{@{}l@{}}Attempted Murder 1st Degree (F,1), \\ Resist/obstruct Officer W/viol (F,1), \\ Agg Battery Grt/Bod/Harm (F,1), \\ Carrying Concealed Firearm (F,1)\end{tabular}        & \begin{tabular}[c]{@{}l@{}}Armed Sex Batt/vict \\ 12 Yrs + (F,2), Aggravated \\ Assault W/dead Weap (F,3), \\ Kidnapping (F,1)\end{tabular} \\ \hline
\begin{tabular}[c]{@{}l@{}}Fernando \\ Walker\end{tabular}   & 1                                                                  & 22                                                  & \begin{tabular}[c]{@{}l@{}}Aggrav Battery w/Deadly Weapon (F,1), \\ Driving Under The Influence (M,2), \\ Carrying Concealed Firearm (F,1)\end{tabular}                                           &                                                                                                                                             \\ \hline
\begin{tabular}[c]{@{}l@{}}Steven \\ Glover\end{tabular}     & 1                                                                  & 28                                                  & \begin{tabular}[c]{@{}l@{}}Robbery / Deadly Weapon (F,11), \\ Poss Firearm Commission Felony (F,7)\end{tabular}                                                                                   &                                                                                                                                             \\ \hline
\begin{tabular}[c]{@{}l@{}}Rufus \\ Jackson\end{tabular}     & 1                                                                  & 40                                                  & \begin{tabular}[c]{@{}l@{}}Resist/obstruct Officer W/viol (F,3), \\ Battery on Law Enforc Officer (F,2), \\ Attempted Robbery Deadly Weapo (F,1), \\ Robbery 1 / Deadly Weapon (F,1)\end{tabular} &                                                                                                                                             \\ \hline
\begin{tabular}[c]{@{}l@{}}Miguel \\ Gonzalez\end{tabular}   & 2                                                                  & 6                                                   & \begin{tabular}[c]{@{}l@{}}Murder in the First Degree (F,1), \\ Aggrav Battery w/Deadly Weapon (F,1), \\ Carrying Concealed Firearm (F,1)\end{tabular}                                            &                                                                                                                                             \\ \hline
\begin{tabular}[c]{@{}l@{}}William \\ Kelly\end{tabular}     & 2                                                                  & 17                                                  & \begin{tabular}[c]{@{}l@{}}Aggravated Assault (F,5), \\ Aggravated Assault W/dead Weap (F,2), \\ Shoot/throw Into Vehicle (F,2), \\ Battery Upon Detainee (F,1)\end{tabular}                      &                                                                                                                                             \\ \hline
\begin{tabular}[c]{@{}l@{}}Richard \\ Campbell\end{tabular}  & 2                                                                  & 21                                                  & \begin{tabular}[c]{@{}l@{}}Armed Trafficking In Cocaine (F,1), \\ Poss Weapon Commission Felony (F,1), \\ Carrying Concealed Firearm (F,1)\end{tabular}                                           &                                                                                                                                             \\ \hline
\begin{tabular}[c]{@{}l@{}}John \\ Coleman\end{tabular}      & 2                                                                  & 25                                                  & \begin{tabular}[c]{@{}l@{}}Attempt Murder in the First Degree (F,1), \\ Carrying Concealed Firearm (F,1), \\ Felon in Pos of Firearm or Amm (F,1)\end{tabular}                                    &                                                                                                                                             \\ \hline
\begin{tabular}[c]{@{}l@{}}Oscar \\ Pope\end{tabular}        & 2                                                                  & 38                                                  & \begin{tabular}[c]{@{}l@{}}Aggravated Battery (F,3), \\ Robbery / Deadly Weapon (F,3), \\ Kidnapping (F,1), \\ Carrying Concealed Firearm (F,2)\end{tabular}                                      & \begin{tabular}[c]{@{}l@{}}Grand Theft in the \\ 3rd Degree (F,3)\end{tabular}                                                              \\ \hline
\begin{tabular}[c]{@{}l@{}}Travis \\ Spencer\end{tabular}    & 3                                                                  & 16                                                  & \begin{tabular}[c]{@{}l@{}}Aggravated Assault W/dead Weap (F,1), \\ Burglary Damage Property\textgreater{}\$1000 (F,1), \\ Burglary Unoccupied Dwelling (F,1)\end{tabular}                        &                                                                                                                                             \\ \hline
\begin{tabular}[c]{@{}l@{}}Michael \\ Avila\end{tabular}     & 3                                                                  & 17                                                  & \begin{tabular}[c]{@{}l@{}}Aggravated Assault W/dead Weap (F,2), \\ Aggravated Assault w/Firearm (F,2), \\ Discharge Firearm From Vehicle (F,1), \\ Home Invasion Robbery (F,1)\end{tabular}      & \begin{tabular}[c]{@{}l@{}}Fail Register \\ Vehicle (M,2)\end{tabular}                                                                      \\ \hline
\begin{tabular}[c]{@{}l@{}}Terrance \\ Murphy\end{tabular}   & 3                                                                  & 20                                                  & \begin{tabular}[c]{@{}l@{}}Solicit to Commit Armed Robbery (F,1), \\ Armed False Imprisonment (F,1), \\ Home Invasion Robbery (F,1)\end{tabular}                                                  & \begin{tabular}[c]{@{}l@{}}Driving While \\ License Revoked (F,3)\end{tabular}                                                              \\ \hline
\begin{tabular}[c]{@{}l@{}}Anthony \\ Hawthorne\end{tabular} & 3                                                                  & 25                                                  & \begin{tabular}[c]{@{}l@{}}Attempt Sexual Batt / Vict 12+ (F,1), \\ Resist/obstruct Officer W/viol (F,1), \\ Poss Firearm W/alter/remov Id\# (F,1)\end{tabular}                                   &                                                                                                                                             \\ \hline
\begin{tabular}[c]{@{}l@{}}Stephen \\ Brown\end{tabular}     & 3                                                                  & 36                                                  & \begin{tabular}[c]{@{}l@{}}Carrying Concealed Firearm (F,2), \\ Battery On Law Enforce Officer (F,1), \\ Kidnapping (F,1), \\ Aggravated Battery (F,1)\end{tabular}                               & \begin{tabular}[c]{@{}l@{}}Driving While \\ License Revoked (F,3)\end{tabular}                                                              \\ \hline
\begin{tabular}[c]{@{}l@{}}Samuel \\ Walker\end{tabular}     & 3                                                                  & 36                                                  & \begin{tabular}[c]{@{}l@{}}Murder in the First Degree (F,1), \\ Poss Firearm Commission Felony (F,1), \\ Solicit to Commit Armed Robbery (F,1)\end{tabular}                                       & \begin{tabular}[c]{@{}l@{}}Petit Theft $100- $300 \\ (M,1)\end{tabular}                                                                     \\ \hline
\begin{tabular}[c]{@{}l@{}}Jesse \\ Bernstein\end{tabular}   & 4                                                                  & 10                                                  & \begin{tabular}[c]{@{}l@{}}Aggravated Battery / Pregnant (F,1), \\ Sex Battery Vict Mental Defect (F,1), \\ Shoot/throw In Occupied Dwell (F,1)\end{tabular}                                      & \begin{tabular}[c]{@{}l@{}}Tresspass in Struct/Convey \\ Occupy (M,1)\end{tabular}                                                          \\ \hline
\begin{tabular}[c]{@{}l@{}}Shandedra \\ Hardy\end{tabular}   & 4                                                                  & 16                                                  & \begin{tabular}[c]{@{}l@{}}Aggrav Battery w/Deadly Weapon (F,1), \\ Felon in Pos of Firearm or Amm (F,4)\end{tabular}                                                                             & \begin{tabular}[c]{@{}l@{}}Resist/Obstruct W/O \\ Violence (M,1), Possess \\ Drug Paraphernalia (M,1)\end{tabular}               
\\ \hline

\end{longtable}
\end{center}
\end{footnotesize}
%\twocolumn